\documentclass[fleqn,usenatbib]{mnras}

\usepackage{newtxtext,newtxmath}

\usepackage[T1]{fontenc}

\DeclareRobustCommand{\VAN}[3]{#2}
\let\VANthebibliography\thebibliography
\def\thebibliography{\DeclareRobustCommand{\VAN}[3]{##3}\VANthebibliography}

\usepackage{graphicx}	
\usepackage{amsmath}	

\title[]{Anisotropic Energy Injection from Magnetar Central Engines in Short GRBs}

\author[]{
Yihan Wang,$^{1,2}$\thanks{yihan.wang@unlv.edu}
Bing Zhang,$^{1,2}$
and Zhaohuan Zhu$^{1,2}$\\
$^{1}$Nevada Center for Astrophysics, University of Nevada, Las Vegas, NV 89154\\
$^{2}$Department of Physics and Astronomy, University of Nevada Las Vegas, Las Vegas, NV 89154, USA\\
}

\date{Accepted XXX. Received YYY; in original form ZZZ}

\pubyear{2015}

\begin{document}
\label{firstpage}
\pagerange{\pageref{firstpage}--\pageref{lastpage}}
\maketitle

\begin{abstract}
A long-lived magnetar, potentially originating from a binary neutron star system, has been proposed to explain the extended emission observed in certain short-duration gamma-ray bursts (sGRBs), and is posited as a potential central engine to power the engine-fed kilonovae. Previously, the process by which energy is injected into the surrounding ejecta/jet was widely believed to be nearly isotropic. In this study, we employ special relativity magnetohydrodynamic (SRMHD) simulations to investigate the wind injection process from a magnetar central engine. We explore the dynamics and energy distribution within the system and found that the parameter $\alpha=u_{\rm A}/u_{\rm MWN}$ can be used to indicate the collimation of the magnetar wind energy injection, where $u_{\rm A}$ is the local Alfven four-speed and $u_{\rm MWN}$ is the four-speed of the magnetar wind nebular (MWN) formed from wind-ejecta collision. A significant portion of the injected energy from the magnetar spin-down wind will be channeled to the jet axis due to collimation within the MWN. Achieving isotropic energy injection requires a significantly small $\alpha$ that necessitates either an ultra-relativistic expanding MWN or an extremely low magnetization MWN, both of which are challenging to attain in sGRBs. Consequently, a considerably reduced energy budget {  (i.e. energy per solid angle reduced by a factor of up to 10 with respect to the value under isotropic assumption}) is anticipated to be injected into the ejecta for engine-fed kilonovae. Engine-fed kilonovae would appear fainter than originally anticipated. 
\end{abstract}




\section{Introduction}
Gamma-ray bursts (GRBs) remain among the most energetic and enigmatic phenomena in the cosmos. Characterized by their intense gamma-ray radiation bursts, GRBs fall into two principal categories: long-duration GRBs (lGRBs) and short-duration GRBs (sGRBs), categorized according to their temporal profiles. While lGRBs correlate with the collapse of massive stars \citep{Woosley1993,Uryu2000, Galama1998, MacFadyen1999, Wheeler2000, Thompson2004}, it is believed that sGRBs emanate from the mergers of binary neutron star (BNS) systems or neutron star-black hole (NS-BH) binaries \citep{Blinnikov1984,Paczynski1986, Eichler1989, Paczynski1991,Narayan1992}. The separation line between the two categories is fuzzy (see recent examples \citealt{zhangbb2021,ahumada2021,rastinejad2022,yang2022,troja2022,levan2023,sun2023,yang2023}), calling for a physical classification scheme based on multi-wavelength criteria \citep{zhang2009}. \\

In recent years, the detection of gravitational waves (GWs) from BNS mergers by terrestrial interferometers \citep{Abbott2017} has furnished persuasive evidence linking sGRBs to these cosmic collisions. Concurrent detection of electromagnetic counterparts, including the optical/IR counterpart known as a kilonova or macronova \citep{Li1998, Metzger2010, Chomiuk2021}, further solidified the association between BNS mergers and sGRBs. It is generally anticipated that emission from a kilonova arises from $r$-process and radioactive decay of heavy elements synthesized in the merger ejecta.\\

Electromagnetic signatures of the sGRB heavily depend on the characteristics of the residual remnant from the mergers. Numerous numerical and theoretical studies \citep{Ruffert1999, Uryu2000, Rosswog2000, Rosswog2003, Oechslin2006, Chawla2010, Rezzolla2011, Hotokezaka2011, Kumar2015, Gao2016, Lasky2016} indicate that the merger product is initially a hypermassive NS (HMNS), maintaining a meta-stable state due to the balance between gravitational collapse and support from differential rotations \citep{Baumgarte2000, Morrison2004, Baiotti2008, O'Connor2011, Lehner2012, Paschalidis2012, Kaplan2014}. Depending on the equation of state (EoS) of the NS, this HMNS can subsequently collapse to form a black hole or persist as an NS. A softer EoS increases the likelihood of the hypermassive NS collapsing to form a black hole, surrounded by an accretion disk and a relatively thick torus of mass $\sim 10^{-2}M_\odot$, the accretion process of which powers the relativistic jet responsible for the GRB \citep{Narayan1992, Rezzolla2011}. Conversely, if the EoS is stiffer, a long-lived supramassive NS (SMNS) or a stable NS may form \citep{Dai2006,Shibata2006,Zhang2013, Giacomazzo2013, Ciolfi2017, Radice2018}. The recent discovery of a $2M_\odot$  NS \citep{Demorest2010, Antoniadis2013} supports the possibility of a stiff EoS, suggesting that the formation of a massive NS is plausible \citep{Ozel2010, Bucciantini2012, Giacomazzo2013, Kiziltan2013}.\\

Once a stable NS is formed, the large angular momentum of the progenitor BNS would cause the newly formed NS to rotate rapidly, necessitating a period approximating one millisecond. Concurrently, the NS would acquire a robust magnetic field as large as approximately $\sim (10^{14} - 10^{15})$ G due to the shear instability at the merger interface \citep{Price2006, Zrake2013}, an $\alpha-\Omega$ dynamo \citep{Duncan1992} or the magnetorotational instability \citep{Akiyama2003, Thompson2005}. This high-spin NS with an extraordinarily strong magnetic field (magnetar) incites characteristic spindown through the emission of magnetic dipole radiation and gravitational radiation \citep{Shapiro1983, Zhang2001}. The spin-down process, absent in the event of prompt BH formation, liberates a significant quantity of energy, proposed as a potential engine for the observed extended emission succeeding the prompt emission of some sGRBs \citep{Zhang2001, Zhang2006, Metzger2008, Perley2009, Troja2008, Lyons2010, Rowlinson2010, Lv2014, Lv2015, Piro2019, Xue2019}. This extended emission can persist for hours, exceeding the timescale predicted by the standard afterglow \citep{Gompertz2013, Rowlinson2013}. While alternative mechanisms like 'fall back' accretion may also power the late emission \citep{Rosswog2007, Lee2009,Cannizzo2011, Kisaka2015, Gibson2017,Lu2023}, whether these proposed models can account for the rich sGRB extended emission and internal plateau phenomenology remains unclear. \\

Assuming a long-lived, strongly magnetized NS can form post a BNS merger, the outflow from the magnetar wind collides with the expanding ejecta, producing a hot proto-magnetar nebula \citep{Dai2004,Metzger2008,Zhang2013,Yu2013,Metzger2014}. As the toroidal magnetic field accumulates in the nebula, the magnetic forces redirect the equatorial outflow towards the pole \citep{Bucciantini2012}. At the initial stage, the Poynting flux launched by magnetar spin-down or differential rotation has a high luminosity, and the collimated outflow could eventually break out of the ejecta, forming a relativistic bipolar jet. Later, a steady Poynting flux  continuously injects into the ejecta, heating the ejecta and the nebula behind it \citep{Gao2013, Yu2013, Metzger2014}. If the ejecta expands sufficiently for photons to diffuse outward on the expansion timescale, the emission from the nebula could escape to the observer. Coupled with the emission powered by the radioactive decay of heavy nuclei synthesized in the ejecta, a more luminous 'kilonovae' could be observed \citep{Yu2013, Metzger2014, Yu2018, Li2018, Ai2022}, which could be possible  promising evidence for a long-lived magnetar remnant. However, previous studies on such kind of engine-fed kilonovae, which assumed an almost isotropic magnetar wind injection to ejecta with a sufficiently large heating efficiency, ignored the potential wind collimation and detailed heating process \citep{Yu2013,Metzger2014,Wollaeger2019,Wang2023}. \cite{Ai2022} found that the injection-induced heating is efficient only during the shock crossing phase, which occurs right after the magnetar wind collides with the ejecta. After the ejecta has been shocked, there is a significant drop in heating efficiency. Given that the shock crossing timescale is relatively short compared to the magnetar spin-down timescale, a low overall heating efficiency is expected.
Whether enough energy could be employed to heat the ejecta remains questionable.\\

Although there remain challenges numerically for sGRBs powered by a magnetar central engine (but see \citealt{Most2023a,Most2023b} for recent encouraging development), recent X-ray observations of GRB 230307A by LEIA provide a strong indication of the existence of such a magnetar central engine \citep{sun2023}. Understanding the efficiency of energy injection from the magnetar wind into the surrounding ejecta/jet is still lacking, specifically concerning the angular distribution of the injected energy {  from magnetar spin-down} and its dependence on various factors. To address these knowledge gaps, we undertake a comprehensive study of the wind injection process from a magnetar in sGRBs through detailed numerical simulations based on the framework of special relativistic magnetohydrodynamics (SRMHD). we aim to clarify the efficiency and angular dependency of energy injection into the surrounding medium, thereby shedding light on the mechanisms driving the extended emission and kilonova emission in sGRBs.\\

Section 2 of this paper details the physical models, numerical methods, and initial conditions used in our simulations, including the boundary conditions necessary for accurately capturing the complex interactions between the magnetar wind and its surroundings. Section 3 provides an analysis of the simulation results, focusing on the efficiency of energy injection and its angular distribution. We explore the physical processes leading to the preferential deposition of energy into specific regions of the outflow, such as the poles. The implications of our findings are discussed in Section 4, where we compare our results with observational data and previous theoretical studies. We evaluate the implications of the angular dependency of energy injection for the observed features and variability of extended emission in sGRBs. Furthermore, we discuss our findings' implications for understanding magnetar wind dynamics and their connection to the overall energetics and temporal behavior of sGRB emissions.

\section{Methods}
{ In our numerical simulations, we first model the injection of an artificial top hat jet into the dynamical ejecta resulting from a BNS merger. This is followed by the simulation of magnetar wind injection, which emanates from the spin-down of a magnetar formed after the merger. The wind injection process commences only after the cessation of the jet's power from the central engine.\\
}
All the simulations in this paper use the relativistic MHD code {\tt Athena++} with HLLD Riemann solver \citep{Stone2020}. The relativistic MHD equations are written as
\begin{eqnarray}
\frac{\partial D}{\partial t} + \nabla\cdot(D\mathbf{v})=0\,,\\
\frac{\partial \mathbf{{M}}}{\partial t} + \nabla\cdot\mathbf{{S}}=0\,,\\
{ \frac{\partial \mathcal{E}}{\partial t} + \nabla\cdot\mathbf{{M}}=0}\,,\\
\frac{\partial \mathbf{B}}{\partial t} - \nabla\times(\mathbf{v}\times\mathbf{B})=0\,,
\end{eqnarray}
where $D$, $\mathbf{v}$, $\mathbf{{M}}$, $\mathbf{{S}}$, $\mathcal{E}$ and $\mathbf{B}$ are the density, velocity, momentum { density}, stress tensor, total energy { density} and magnetic field strength in the lab frame, respectively. $p_g$, $\rho$ and $\Gamma$ are gas pressure, density, and adiabatic index in the co-moving frame.
The density, momentum density and strees tensor can be expressed as
\begin{eqnarray}
D&=&\gamma \rho \,,\\
\mathbf{{M}}&=&({ \mathcal{E}}+p_g+p_m)\mathbf{v}-(\mathbf{v}\cdot\mathbf{B})\mathbf{B}\,,\\
\mathbf{{S}}&=&\gamma^2w\mathbf{v}\mathbf{v}-\frac{1}{\gamma^2}\mathbf{B}\mathbf{B}-(\mathbf{v}\cdot\mathbf{B})(\mathbf{v}\mathbf{B}+\mathbf{B}\mathbf{v})\nonumber\\
&-&\gamma^2(\mathbf{v}\cdot\mathbf{B})^2\mathbf{v}\mathbf{v}+(p_g+p_m)\mathbf{I}\,,\\
p_m&=&\frac{1}{2}\left(\frac{1}{\gamma^2}\mathbf{B}^2+(\mathbf{v}\cdot\mathbf{B})^2\right)\,,\\
w&=&\rho+\frac{\Gamma}{\Gamma-1}p_g+2p_m\,, \\
\mathcal{E} &=&\gamma^2 w - (p_g+p_m) -\gamma^2(\mathbf{v}\cdot\mathbf{B})^2\,.
\end{eqnarray}
Viscosity and non-ideal MHD effects that depend on the ionization, reconnection, and collision rate are not considered in our simulations.

The computational domain is constructed in 2D with spherical polar coordinates with 512 cells logarithmically in the radial direction ranging from $r_{\rm in}=4.7\times10^{8}$ cm to $r_{\rm out}=9.6\times10^{11}$ cm and 512 cells uniformly in the polar angle direction ranging from 0 to $\frac{\pi}{2}$. The reflective boundary conditions are set on the polar axis and equatorial axis. Outflow boundary conditions are set on the outer boundary of the radial direction. The inner boundary conditions of the radial direction are time-dependent. If there is no injection, the boundary conditions are set to be reflective to guarantee zero energy injection. If there are energy/matter injections, e.g. jet or wind, the boundary conditions are set accordingly based on the jet/wind models that will be discussed as follows. A { gamma-law} EoS is adopted for all the simulations with $\Gamma = 4/3$.

\subsection{Boundary conditions and initial conditions for sGRB}
\subsubsection{Ejecta models}\label{sec:bc-ej}
Our simulations start at $\sim 0.5$ s after the BNS merge. Therefore, an ejecta profile is initially set up in the computational domain.  The typical velocity of the ejecta is $v_{\rm ej}=$0.25 c \citep{kasliwal2017}, after 0.5 s of the BNS merger, the ejecta has propagated to a typical radius $r_{\rm ej}\sim 3.75\times10^9$ cm. We assume the ejecta expand adiabatically after the BNS merger, thus homogenous density and velocity profiles can be used for the initial conditions in our simulations, 
\begin{eqnarray}
\rho(r, \theta)&=&\rho_{\rm ej}(\theta)\frac{r_{\rm ej}^2}{r^2},\quad r_{\rm in}\le r \le r_{\rm ej}\,,\\
\mathbf{u}(r) &=& u_{\rm ej}\frac{r}{r_{\rm ej}}\hat{\mathbf{r}},\quad r_{\rm in}\le r \le r_{\rm ej},
\end{eqnarray}
where $\mathbf{u}$ is the four-velocity and $u_{\rm ej}$ is the corresponding four-velocity of $v_{\rm ej}$\footnote{Note that the velocity of the ejecta at the inner boundary $r_{\rm in}$ $\sim$ 0.025 c is larger than the local escape velocity $\sim$ 0.015 c, although the gravity of the merged compact object is ignored in our simulations.}. The ejecta density is integrated to be the total mass of the ejecta $\int_{-1}^1d\cos(\theta)\int_0^{2\pi}d\phi\int_{r_{\rm in}}^{r_{\rm ej}} dr \rho(r,\theta) = M_{\rm ej}$. GRMHD simulations indicate that the ejected mass from BNS mergers ranges from $10^{-4}M_\odot$ to $10^{-2}M_\odot$ \citep{Hotokezaka2013,shibata2017}. However, higher ejecta mass is also allowed. To cover the most of parameter space that we are interested in, we adopt three total ejecta masses $10^{-1}M_\odot$, $10^{-2}M_\odot$ and $10^{-3}M_\odot$. The ejecta is assumed to be initially cold with $\frac{p}{\rho c^2}=10^{-5}.$ 

To test the effects of the ejecta profile in the $\theta$ direction on wind propagation, we use two different density profiles, an experimental uniform profile,
\begin{equation}
\rho_{\rm ej}(\theta) = {\rm const},
\end{equation}
where ejected materials from BNS mergers are isotropically distributed in space, and a more practical structure profile
\begin{equation}
\rho_{\rm ej}(\theta) \propto \frac{1}{4}+\sin(\theta)^3,
\end{equation}
where matters are ejected more in the equatorial direction than the polar direction, the density in the equatorial plane is 5 times larger than the density in the pole.

\subsubsection{Jet model}
The jet launching mechanism is still under debate for BNS mergers, although several theories have been proposed. Accretion onto BH/NS and differential rotation of an NS are the two main ways of launching a jet from a magnetar.  Erratic magnetic bubble injection during an HMNS phase supported by differential rotation is another possibility.  Regardless of the central engine and detailed launching mechanism, we inject a top-hat jet during the first 1 second of the simulation near the pole within an openning angle of $\theta_j=10^\circ$.

The acceleration of the short GRB jet can be driven by two mechanisms: thermal acceleration driven by neutrino heating from the accretion disk or proto-neutron star \citep{Meszaros1993, Piran1999} or magnetic acceleration driven by the Poynting-flux launched from the central BH or neutron star \citep{Granot2011}. Realistically, the GRB jet composition is likely ``hybrid'' that includes both components \citep{Gao2015}. The former is relevant for a hot fireball that accelerates the jet rapidly while the latter is relevant for a Poynting-flux dominated outflow that accelerates the jet rapidly to fast magnetosonic speed ($\gamma\sim(1+\sigma_{\rm j,0})^{1/3}$, where $\sigma_{\rm j, 0}$ is the magnetization of the jet in the inner boundary { defined below in Equation~\ref{eq:sigma_j}}) and then keeps slowly accelerating the jet to the terminal Lorentz factor at the coasting radius due to significant magnetic dissipation or impulsive acceleration due to a magnetic pressure gradient within the pulse. In order to include both mechanisms, the jet in our simulations is assumed to be hot and Poynting-flux dominated with isotropic luminosity $L_{\rm j}=5\times10^{51}$ erg/s, terminal Lorentz factor $\gamma_\infty=300$ and initial magnetization $\sigma_{\rm j,0}= 5$\footnote{{ Practically, we do not use a very high $\sigma$ to ensure lower numerical noises. As the value of $\sigma$ becomes extremely high, the gas pressure constitutes only a very small fraction of the total pressure, which includes both gas and magnetic pressures. In most MHD (magnetohydrodynamics) codes, the gas pressure must be derived from the total pressure based on the magnetic field calculated at each step. A higher $\sigma$ results in a lower proportion of gas pressure within the total pressure, leading to significant numerical errors when deriving the gas pressure. Extremely high values of $\sigma$ may also cause negative gas pressure due to numerical inconsistencies between the calculated total pressure and magnetic field strength.}}. The boundary conditions\footnote{Note that $\rho$ and $p$ are written in the co-moving frame and $B$ is written in the lab frame.} for the jet injection ($\theta<\theta_j$) are \citep{Geng2019}
\begin{eqnarray}
v_{\rm j,r} &=& 0.8 c\,,\\
\rho_{\rm j} &=& \frac{L_{\rm j}}{4\pi r_{\rm in}^2v_{\rm j,r}\gamma_0^2hc^2}\,,\\
p_{\rm j} &=& \rho_{\rm j}c^2(\eta-1)\frac{\Gamma-1}{\Gamma}\,,\\
B_{\rm j,\phi}^2 &=& \gamma_0^2\rho_{\rm j}c^2 \eta\sigma_{\rm j},\quad B_{\rm j,\theta} = 0,\quad B_{\rm j,r} =0, \label{eq:sigma_j}
\end{eqnarray}
where $\gamma_0$ is the initial Lorentz factor, { $\eta=(\rho_{\rm j}c^2+\frac{\Gamma-1}{\Gamma}p_{\rm j})/(\rho_{\rm j}c^2)$ is the specific thermal enthalpy}, and $h=(\rho_{\rm j}c^2+\frac{\Gamma-1}{\Gamma}p_{\rm j}+\frac{B_{\rm j, \phi}^2}{\gamma_0^2})/(\rho_{\rm j}c^2)=\eta(1+\sigma_{\rm j})=\gamma_\infty/\gamma_0$ is the specific total enthalpy.

\subsubsection{Wind models}\label{sec:wind}

\begin{figure}
\includegraphics[width=1\columnwidth]{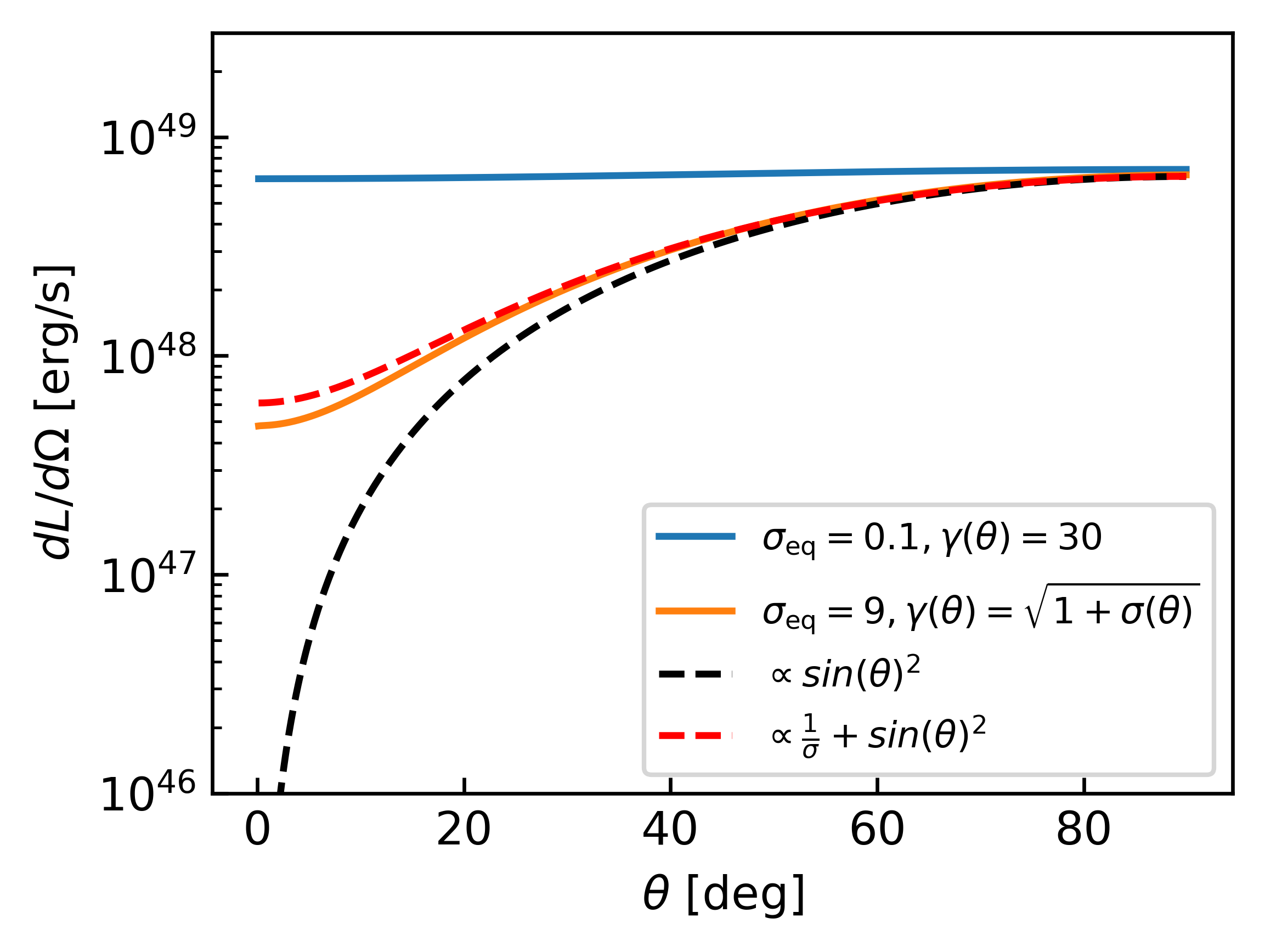}
\caption{Luminosity profile for matter-dominated wind (blue line) and Poynting-flux-dominated wind (orange line) at the injection boundary. A significantly higher monopole-like spin-down luminosity $\sim 10^{49}$ erg/s than the dipole spin-down luminosity $\sim 10^{47}$ erg/s of a magnetar is adopted in our simulations. }
\label{fig:wind-bc}
\end{figure}

If the total mass of the binary NS is low or the equation of state of the NS is stiff enough, the BNS merger remnant could be a rapidly spinning SMNS or a stable NS rather than a black hole. Due to the rapid spin, it is plausible that the proto-NS will develop a strong magnetic field by $\alpha$-$\Omega$ dynamo \citep{Duncan1992} or by shear/magnetorotational  instability \citep{Price2006, Akiyama2003, Thompson2005}. 
The proto-NS initially loses mass through a neutrino-driven wind. This wind has high baryon loading and therefore is dirty (non-relativistic). The flow becomes clean after the NS cools down when the neutrino luminosity drops \citep{Qian1996, Metzger2011}.
Once the meta-stable proto-NS becomes stable by losing sufficient mass from the outflow, a long-lived millisecond magnetar can be formed. The newly formed magnetar can serve as the engine of the extended emission of the sGRB or engine-fed kilonova with its spin-down radiation. For an orthogonal rotating magnetar with a dipole magnetic field, the spin-down luminosity is
\begin{eqnarray}
L_{\rm sd}&=&\frac{2}{3c^3}B_\star^2R_\star^6\bigg(\frac{2\pi}{P}\bigg)^4\\
&\sim&10^{47}\bigg(\frac{B_\star}{10^{14} \rm G}\bigg)^2\bigg(\frac{R_\star}{10^6 \rm cm}\bigg)^6\bigg(\frac{P}{1 \rm ms}\bigg)^{-4} \rm erg/s \,,
\end{eqnarray}
associated with a corresponding spin-down timescale 
\begin{equation}
t_{\rm sd}\sim 10^{5}\bigg(\frac{B_\star}{10^{14} \rm G}\bigg)^{-2}\bigg(\frac{R_\star}{10^6 \rm cm}\bigg)^{-6}\bigg(\frac{P}{1 \rm ms}\bigg)^2 \rm s \,,\label{eq:sdt}
\end{equation}
where $B_\star$, $R_\star$, and $P$ are the surface magnetic field strength, radius, and period of the star, respectively. 

Simulating the magnetar wind injection with $\sim 10^{5}$ s is impractical. This requires an extremely large computing domain extending to $10^{15}-10^{16}$ cm. To reduce the cost of the simulations so that we can explore more regions of the parameter space, we use an experimental monopole-like magnetic field. The spin-down luminosity of the monopole field is much higher than the dipole field, which scales as
\begin{eqnarray}
L_{\rm sd, mono} &=& \frac{2}{3c}B_\star^2R_\star^4\bigg(\frac{2\pi}{P}\bigg)^2\\
&\sim&10^{49}\bigg(\frac{B_\star}{10^{14} \rm G}\bigg)^2\bigg(\frac{R_\star}{10^6 \rm cm}\bigg)^4\bigg(\frac{P}{1 \rm ms}\bigg)^{-2} \rm erg/s \,.
\end{eqnarray}
Unlike the dipole magnetic field, the $B(r)\propto1/r^3$ within the light cylinder, the monopole-like magnetic field scales with $1/r^2$ within the light cylinder. Therefore, in the monopole-like model, the winding magnetic field (toroidal) that is built by a rotating poloidal field is much stronger than the dipole field model. This allows us to use a reasonable computing domain to carry out the simulations. We use the fiducial parameters $R_\star = 1.2\times10^6$ cm, $B_\star=10^{15}$ G and $P=$5 ms. 

Before the wind collides with the ejecta, if the magnetic acceleration is efficient due to strong magnetic dissipation, the wind can be accelerated to a super-magnetosonic speed, meanwhile, the magnetization drops slowly towards zero. As a result, the wind becomes matter-dominated with an extreme Lorentz factor. If the magnetic acceleration is not efficient, the wind can only be accelerated to the magnetosonic speed. The magnetization parameter will remain high in this case. Thus, the wind will be still Poynting-flux dominated.  We consider both cases in our simulations { by giving different input wind Lorentz factors $\gamma_{\rm w, eq}$ and wind magnetizations $\sigma_{\rm w, eq}$ at the equator. Then the wind toroidal magnetic field $B_{\rm w,\phi}$ and wind density $\rho_w$ at the boundary where $r_{\rm in}=4.7\times 10^8$ cm can be calulated}
{ 
\begin{eqnarray}
B_{\rm w,\phi}^2&=&\frac{3L_{\rm sd, mono}}{2r_{\rm in}^2c}\sin(\theta)^2,\quad B_{\rm w,\theta}=B_{\rm w,r}=0\,,\\
\rho_{\rm w}&=&\frac{B_{\rm w,\phi}^2|_{\theta=\pi/2}}{\gamma_{\rm w,eq}^2c^2\sigma_{\rm w,eq}}\,,
\end{eqnarray}
}
For a super-magnetosonic matter-dominated wind, we use $\gamma_{\rm w} = \gamma_{\rm w, eq}= 30$, and $\sigma_{\rm w, eq}=0.1$. For a magnetosonic Poynting-flux-dominated wind, we use $\gamma_{\rm w}(\theta) \sim u_A/c=\sqrt{1+\sigma_{\rm w}(\theta)}$ and $\sigma_{\rm w, eq}=9$, where $u_A$ is the four velocity of magnetosonic speed. {  The magnetic field strength at the surface of the neutron star $B_\star$ is fixed to be $10^{15}$~G and the neutron star radius $R_\star$ is set to be $10^6$ cm. We have adopted magnetar spinning period $P=5$ ms\footnote{Because we use a split monopole wind, which has a spin-down luminosity almost two orders of magnitude larger than that of the dipole law, the long periods were adopted to compensate for this. This adjustment helps maintain a reasonable spin-down luminosity given the total spin energy. } in our simulations. }

Figure~\ref{fig:wind-bc} shows the angular distribution of the wind luminosity for $P=5$ ms. As indicated, the wind is essentially isotropic for a matter-dominated wind. For a Poynting-flux-dominated wind, on the other hand, the injection luminosity is much higher near the equator.

\subsection{Boundary conditions and initial conditions for test cases}\label{sec:ic-test}
To investigate the isotropy of magnetar wind injection, we conducted experimental simulations of isotropic wind injection into an ideal static isotropic medium with varying medium densities and wind properties, including different magnetizations and Lorentz factors.

The medium is configured to be fully isotropic throughout the computational domain, with zero magnetization and uniform density ($\rho_{\rm amb}=10^{-1},10^{-3}$, and $10^{-5}$ in code unit). Additionally, the medium is assumed to be cold, characterized by a pressure-to-density ratio of $\frac{p_{\rm amb}}{\rho_{\rm amb}c^2}=10^{-3}$.

The wind is injected at the inner radial boundary, exhibiting isotropic density and a uniform toroidal magnetic field in the $\theta$ direction. This wind maintains a constant total luminosity, adapting to various magnetizations denoted by { $\sigma_w=0.01,0.1,1$, and $10$.} The outer radial boundary is set as an outflow boundary, while the two polar boundaries are configured to be reflective.
{ 
\begin{table*}
\begin{tabular}{|l|c|c|c|l|c|l|l|c|}
\hline
model & \multicolumn{1}{l|}{$L_{\rm jet}$ {[}erg/s{]}} & \multicolumn{1}{l|}{$\tau_{\rm jet}$ {[}s{]}} & \multicolumn{1}{l|}{$E_{\rm ej}$ {[}erg{]}} & ejecta profile             & \multicolumn{1}{l|}{$E_{\rm wind}$ {[}erg{]}} & $\sigma_{\rm w}$ & $\gamma_{\rm w}$ & \multicolumn{1}{l|}{$\xi_{\rm therm}$ at equator} \\ \hline
1     & $5\times10^{51}$                               & 1                                             & $1.8\times10^{51}$                          & $\frac{1}{4}+\sin^3\theta$ & $10^{52}$                                     & 0.1              & 30               & $\sim$0.1                                         \\ \hline
2     & $5\times10^{51}$                               & 1                                             & $1.8\times10^{51}$                          & $\frac{1}{4}+\sin^3\theta$ & $10^{52}$                                     & 9                & $\sqrt{10}$      & $\sim$0.03                                        \\ \hline
3     & $5\times10^{51}$                               & 1                                             & $1.8\times10^{53}$                          & $\frac{1}{4}+\sin^3\theta$ & $10^{52}$                                     & 0.1              & 30               & $\sim$0.02                                        \\ \hline
4     & $5\times10^{51}$                               & 1                                             & $1.8\times10^{53}$                          & $\frac{1}{4}+\sin^3\theta$ & $10^{52}$                                     & 9                & $\sqrt{10}$      & $<$0.001                                          \\ \hline
5     & no jet                                         & 1                                             & $1.8\times10^{51}$                          & isotropic                  & $10^{52}$                                     & 0.1              & 30               & 0.3                                               \\ \hline
6     & no jet                                         & 1                                             & $1.8\times10^{51}$                          & isotropic                  & $10^{52}$                                     & 9                & $\sqrt{10}$      & 0.05                                              \\ \hline
7     & no jet                                         & 1                                             & $1.8\times10^{53}$                          & isotropic                  & $10^{52}$                                     & 0.1              & 30               & 0.1                                               \\ \hline
8     & no jet                                         & 1                                             & $1.8\times10^{53}$                          & isotropic                  & $10^{52}$                                     & 9                & $\sqrt{10}$      & $<$0.001                                          \\ \hline
9     & no jet                                         & 1                                             & $1.8\times10^{51}$                          & $\frac{1}{4}+\sin^3\theta$ & $10^{52}$                                     & 0.1              & 30               & 0.2                                               \\ \hline
10    & no jet                                         & 1                                             & $1.8\times10^{51}$                          & $\frac{1}{4}+\sin^3\theta$ & $10^{52}$                                     & 9                & $\sqrt{10}$      & 0.03                                              \\ \hline
11    & no jet                                         & 1                                             & $1.8\times10^{53}$                          & $\frac{1}{4}+\sin^3\theta$ & $10^{52}$                                     & 0.1              & 30               & 0.1                                               \\ \hline
12    & no jet                                         & 1                                             & $1.8\times10^{53}$                          & $\frac{1}{4}+\sin^3\theta$ & $10^{52}$                                     & 9                & $\sqrt{10}$      & $<$0.001                                          \\ \hline
\end{tabular}
\caption{A table summarizing the parameters of the jet, ejecta, and wind in our models. }
\label{tab1}
\end{table*}

Table~\ref{tab1} summarizes all the models simulated in this paper.}
\section{Results}

\subsection{Collimation of magnetar wind}\label{sec:pure-wind}



\begin{figure*}
\includegraphics[width=2\columnwidth]{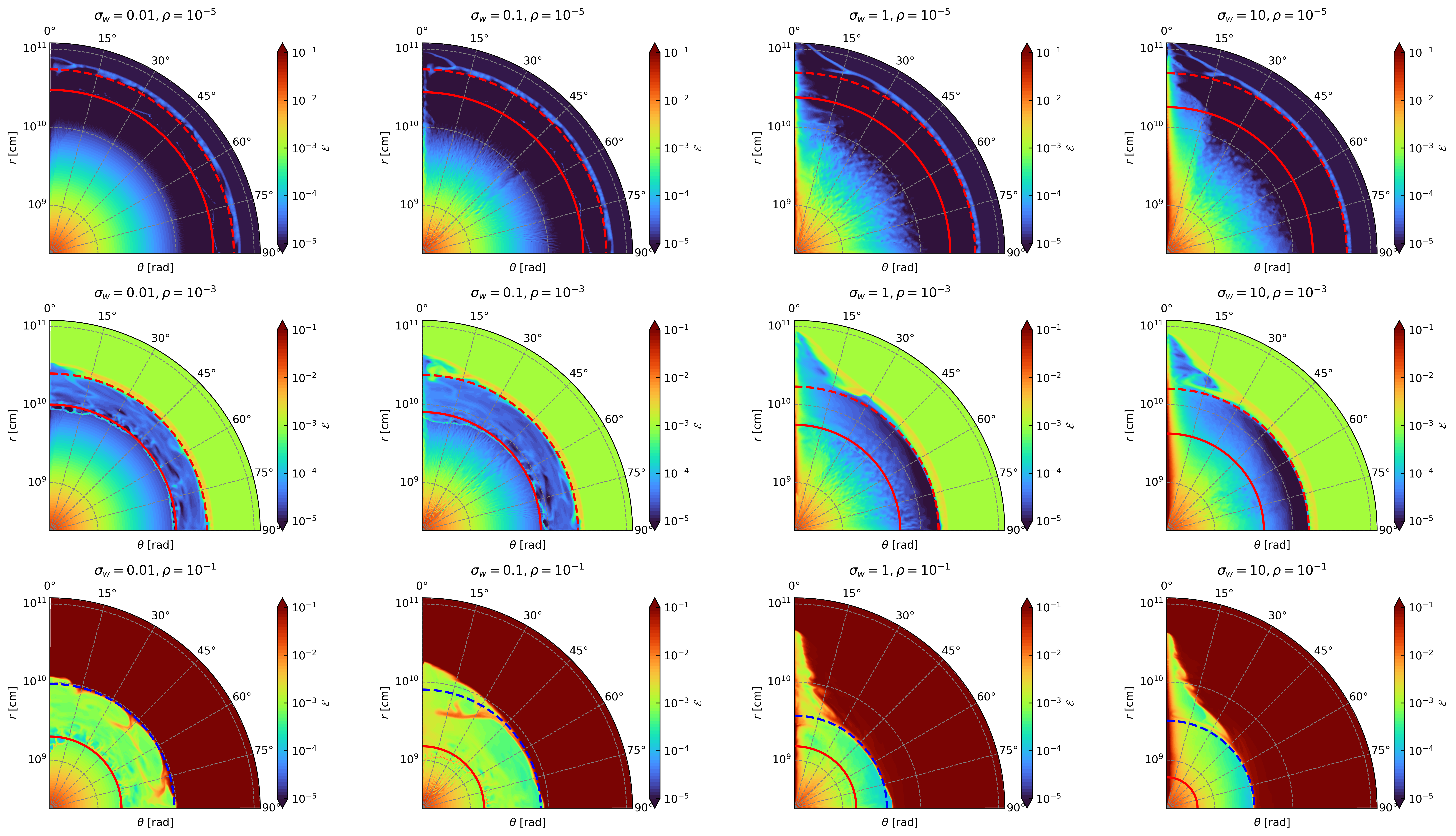}
\caption{Snapshots of energy { density} profile on magnetar wind injection to a static isotropic medium with different injected wind magnetization $\sigma_{\rm w}$ and medium densities. The collimation of the magnetar wind is determined by the magnetization and Lorentz factor of the magnetar wind nebula formed from the wind-medium collision. { Thick dashed lines indicate the contact discontinuities and thick solid lines represent the position of the reverse shocks.} }
\label{fig:wind-amb}
\end{figure*}

{ 

To understand the isotropy of the wind injection under different conditions, we conducted several simulations of wind injection into a uniform ambient medium using the initial and boundary conditions described in Section~\ref{sec:ic-test}, where the wind at the inner boundary is set to be isotropic with a pure toroidal magnetic field and a constant total luminosity.  

Figure~\ref{fig:wind-amb} indicates the energy {  density} profiles 4 seconds after the wind injection. As indicated in the first row, for wind injection into a low-density medium, the wind is isotropic if the magnetization is low $\sigma_{\rm w}=0.01$. As the $\sigma_{\rm w}$ becomes higher and higher, plasma in the MWN starts to focus toward the pole, forming a collimated outflow. As indicated by each column, the collimation also becomes stronger if the density of the medium becomes higher.

Indeed, as the magnetar wind collides with the medium, both the forward and reverse shocks are formed at the contact discontinuity. The reverse shock rapidly propagates inwards and disappears until it reaches the wind accelerating zone \citep{Ai2022}. The reverse shock slows down and heats the magnetar wind, leading to the formation of an MWN. If the medium cannot efficiently slow down the magnetar wind, allowing the wind to propagate nearly freely at relativistic speeds, the hoop stress from the dominant toroidal magnetic field will be effectively balanced by the electric force in the wind, resulting in an asymptotic force-free solution. In this scenario, wind injection is nearly isotropic. However, if the medium can considerably decelerate the magnetar wind, the plasma within the MWN transitions to a non-relativistic state. The electric force that counterbalances the hoop stress, attempting to constrict the magnetized outflow from the strong toroidal magnetic field, becomes weak. In this scenario, if the MWN is highly magnetized, the substantial hoop stress within the MWN begins to focus the magnetar wind towards the polar region, resulting in a collimated outflow. On the other hand, if the MWN is matter-dominated with low magnetization, the MWN will expand more isotropically. Plots illustrating the Coulomb force and Lorentz force under various conditions are included in the Appendix.

To quantify the strength of the collimation and investigate how the strength of the collimation correlates with the magnetization and Lorentz factor of the MWN, we define a parameter
\begin{eqnarray}
\zeta &=& \frac{1}{\pi}\int{\rm max}\left(\frac{L_{\rm iso}(\theta)}{L_{\rm w}},\frac{L_{\rm w}}{L_{\rm iso}(\theta)}\right){\rm d}\theta\\
\frac{L_{\rm iso}(\theta)}{L_{\rm w}} &=& \frac{{\rm d}E/{\rm d}\Omega}{E/{4\pi}}
\end{eqnarray}
to quantify the strength of the MWN outflow collimation, where $L_{\rm w}$ is the injection luminosity of the magnetar wind and $L_{\rm iso}(\theta)$ is the observed isotropic luminosity with viewing angle $\theta$ and $E$ is the total injected energy from the wind. The larger $\zeta$ is, the stronger the collimation will be. $\zeta=1$ represents that the wind is isotropic as it was before the injection. 

We also define a energy weighted parameter $\langle \alpha_{\rm MWN}\rangle_{\rm E}$ to describe the property of the MWN,
\begin{eqnarray}
\langle \alpha_{\rm MWN}\rangle_{\rm E} &=&\frac{\int \alpha \mathcal{E}{\rm d}V }{\int \mathcal{E}{\rm d}V} \,,\\
\alpha &=& {\frac{u_{\rm A}}{u_{\rm MWN}}}\,,
\end{eqnarray}
where $u_{\rm A}$ and $u_{\rm MWN}$ are the local Alfven four-speed and four-velocity of the MWN, respectively. 

\begin{figure}
\includegraphics[width=\columnwidth]{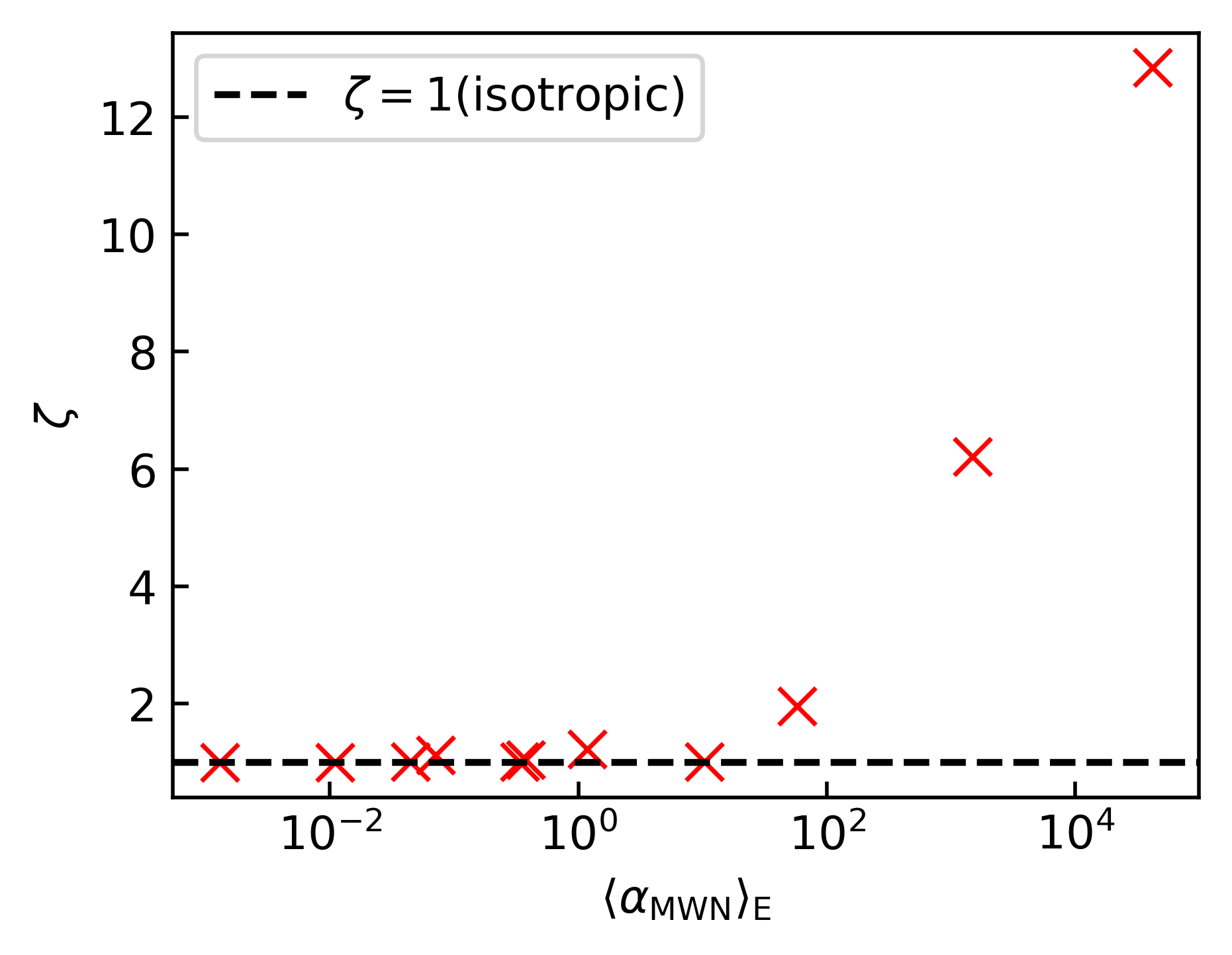}
\caption{Collimation strength as a function of the energy weighted $\alpha$ in the magnetar wind nebula. $\alpha=u_{\rm A}/u_{\rm MWN}$ is a dimensionless parameter that represents the ratio between the Alfven four-speed and the four-velocity of the magnetar wind nebula. }
\label{fig:alpha-xi}
\end{figure}
Figure~\ref{fig:alpha-xi} shows the relationship between $\zeta$ and $\langle \alpha_{\rm MWN}\rangle_{\rm E}$. It is shown that the magnetar wind injection stays isotropic if $\langle \alpha_{\rm MWN}\rangle_{\rm E}$ is below $\sim1$ and the collimation starts to appear if $\langle \alpha_{\rm MWN}\rangle_{\rm E}$ is high enough. In general, a higher $\langle \alpha_{\rm MWN}\rangle_{\rm E}$ would indicate a stronger collimation. Therefore, $\langle \alpha_{\rm MWN}\rangle_{\rm E}$ of the MWN can be used as an indicator for the magnetar wind injection anisotropy.
}

\subsection{Magnetar wind injection in sGRBs}
\subsubsection{Jet injection}

\begin{figure*}
 \includegraphics[width=2\columnwidth]{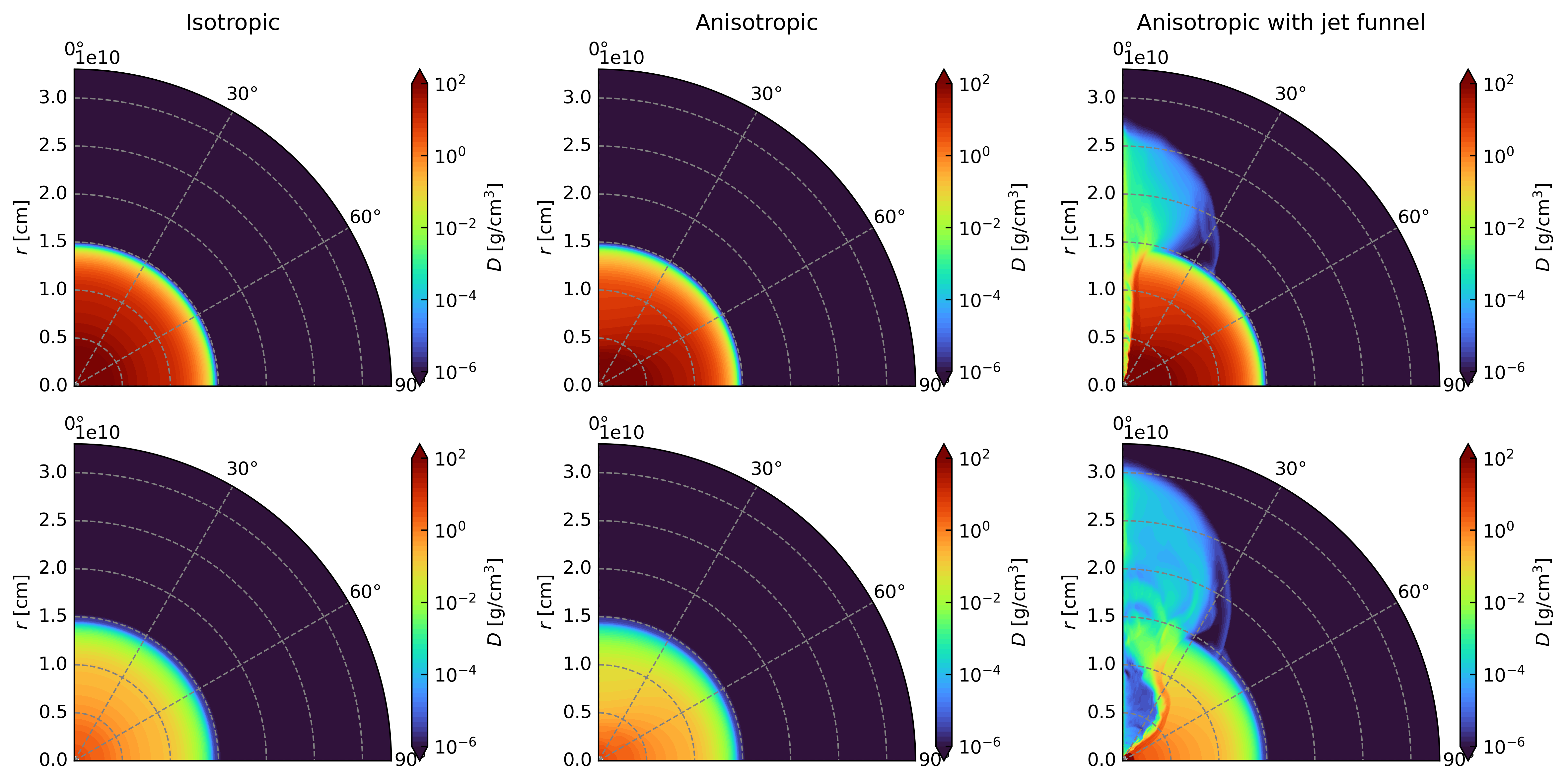}
    \caption{Snapshots of the density profile before magnetar wind injection. Upper panels: $M_{\rm ej}=10^{-1}M_\odot$; Bottom panels: $M_{\rm ej}=10^{-3}M_\odot$; Left panels: isotropic ejecta; Middle panels: anisotropic ejecta; Right panels: anisotropic ejecta with jet injection; The snapshots are taken at t=1 s, right after the jet injection finished.}
    \label{fig:pre-wind-shot}
\end{figure*}

Before the formation of a long-lived magnetar, the differential rotation of the proto-magnetar or the accretion process shortly after the BNS merger may initiate the launch of a jet. If a jet can be produced before the magnetar wind is launched, it can create a funnel within the ejected material. This funnel significantly affects the injection of the wind. Therefore, before initiating the wind launch, we need to investigate the jet injection process. Figure~\ref{fig:pre-wind-shot} displays density snapshots of different ejecta models at $t=1$ s (indicating completion of jet injection if applicable). The upper panels show the ejecta model with a total mass of $M_{\rm ej}=10^{-1}M_\odot$, while the bottom panels show the ejecta model with a smaller total mass of $M_{\rm ej}=10^{-3}M_\odot$. In the last row of this figure, it is evident that if the ejecta mass is sufficiently large (resulting in higher pressure to confine the injected jet), a narrow funnel can be formed. Conversely, if the total mass of the ejecta is small (leading to lower pressure within the ejecta), a very wide funnel will be formed.

\subsubsection{Matter-dominated wind injection}\label{sec:m-wind}
\begin{figure*}
\includegraphics[width=2\columnwidth]{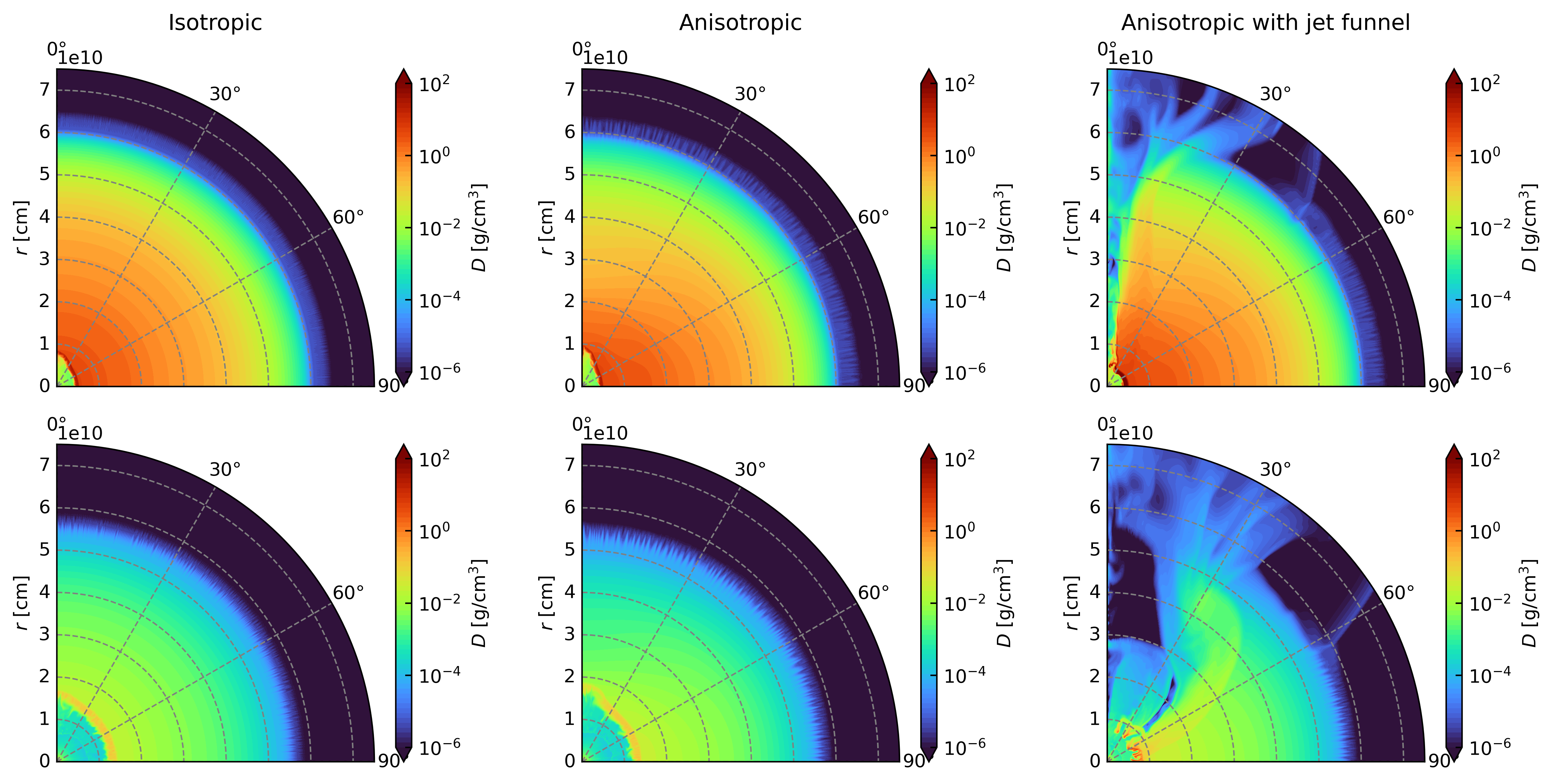}\\
\caption{Snapshots of the density profile after magnetar wind injection (matter-dominated at the injection boundary {with $\sigma=0.1$}). Upper panels: $M_{\rm ej}=10^{-1}M_\odot$ (model 7, 11  and 3); Bottom panels: $M_{\rm ej}=10^{-3}M_\odot$ (model 5, 9, 1); Left panels: isotropic ejecta; Middle panels: anisotropic ejecta; Right panels: anisotropic ejecta with jet injection; The snapshots are taken at t=5 s, including 1 s jet working time, 3 s waiting time and 1 s wind injection time.}
\label{fig:m-wind-rho}
\end{figure*}

Figure~\ref{fig:m-wind-rho} displays density snapshots at {$t=5$} s, which includes 1 s of jet working time, 3 s of waiting time, and 1 s of wind injection time, for matter-dominated {($\sigma_w=0.1$)} wind injections into different types of ejecta. The upper panels represent cases with a total ejecta mass of $10^{-1}M_\odot$, while the lower panels correspond to cases with a total ejecta mass of $10^{-3}M_\odot$. In both cases, the reverse shock that arose from the wind-ejecta collision can significantly reduce the speed of the magnetar wind, thus  $\langle \alpha_{\rm MWN}\rangle_{\rm E}$ becomes large enough for the collimation to appear.

{
In the case where the ejecta profile is non-isotropic as described in Section~\ref{sec:bc-ej}, with a lower density/pressure at the poles, the pressure gradient of the ejecta provides a weak additional collimation effect on the magnetar wind, as shown in the middle columns of this figure. If a pre-existing funnel exists within the ejecta, the wind propagation is significantly altered. At the moment of the magnetar wind colliding with the ejecta, if the plasma in the funnel remains sufficiently hot (the 3-second waiting time is insufficient for efficient cooling), the total pressure within the funnel can still exceed the pressure within the ejecta. This high pressure impedes the flow of the magnetar wind through the funnel, resulting in the concentration of more energy around the funnel boundary. However, as the plasma in the funnel cools down, the pressure within the funnel decreases below the pressure of the expanding ejecta, allowing the magnetar wind to pass freely through the funnel. Consequently, a highly collimated outflow breaks out through the funnel.

The bottom panels of Figure~\ref{fig:m-wind-rho} illustrate cases with a smaller total ejecta mass of $10^{-3}M_\odot$. Due to the lower mass of the ejecta, the MWN expands faster than the cases with an ejecta mass of $10^{-1}M_\odot$. Since the MWN expands faster, $\langle \alpha_{\rm MWN}\rangle_{\rm E}$ becomes smaller. As a result, the collimation is weaker in these cases exhibiting a larger outflow opening angle.}

\begin{figure*}
\includegraphics[width=2\columnwidth]{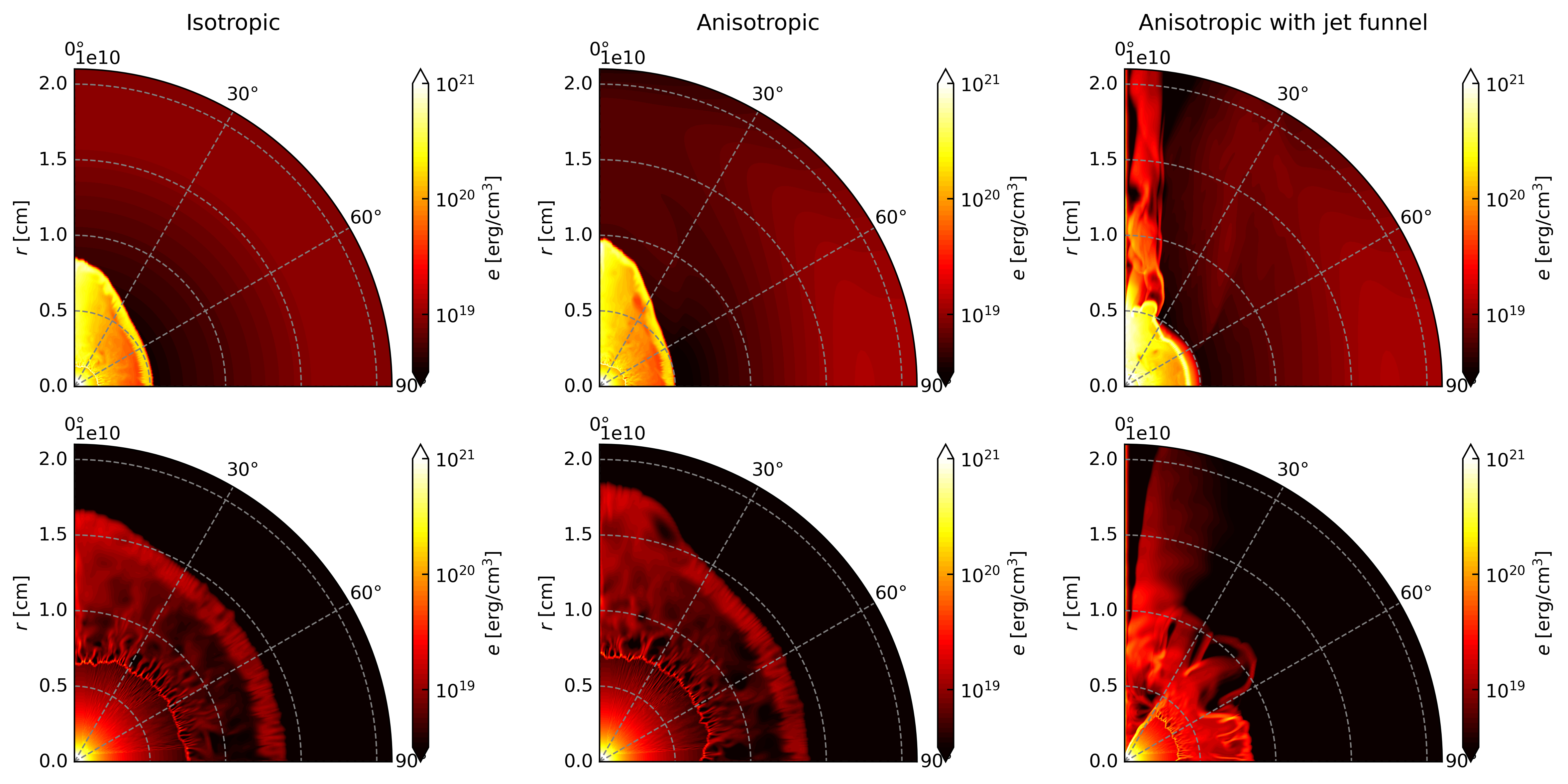}\\
\caption{Similar to Figure~\ref{fig:m-wind-rho}, but for energy density snapshots.}
\label{fig:m-wind-e}
\end{figure*}
Figure~\ref{fig:m-wind-e} shows the corresponding energy density of Figure~\ref{fig:m-wind-rho}. Comparing the middle panels and the left panels, we can clearly see the extra collimation from the pressure/density of the ejecta. The right panels show the energy flow in the funnel.

\subsubsection{Poynting-flux dominated wind injection}

\begin{figure*}
\includegraphics[width=2\columnwidth]{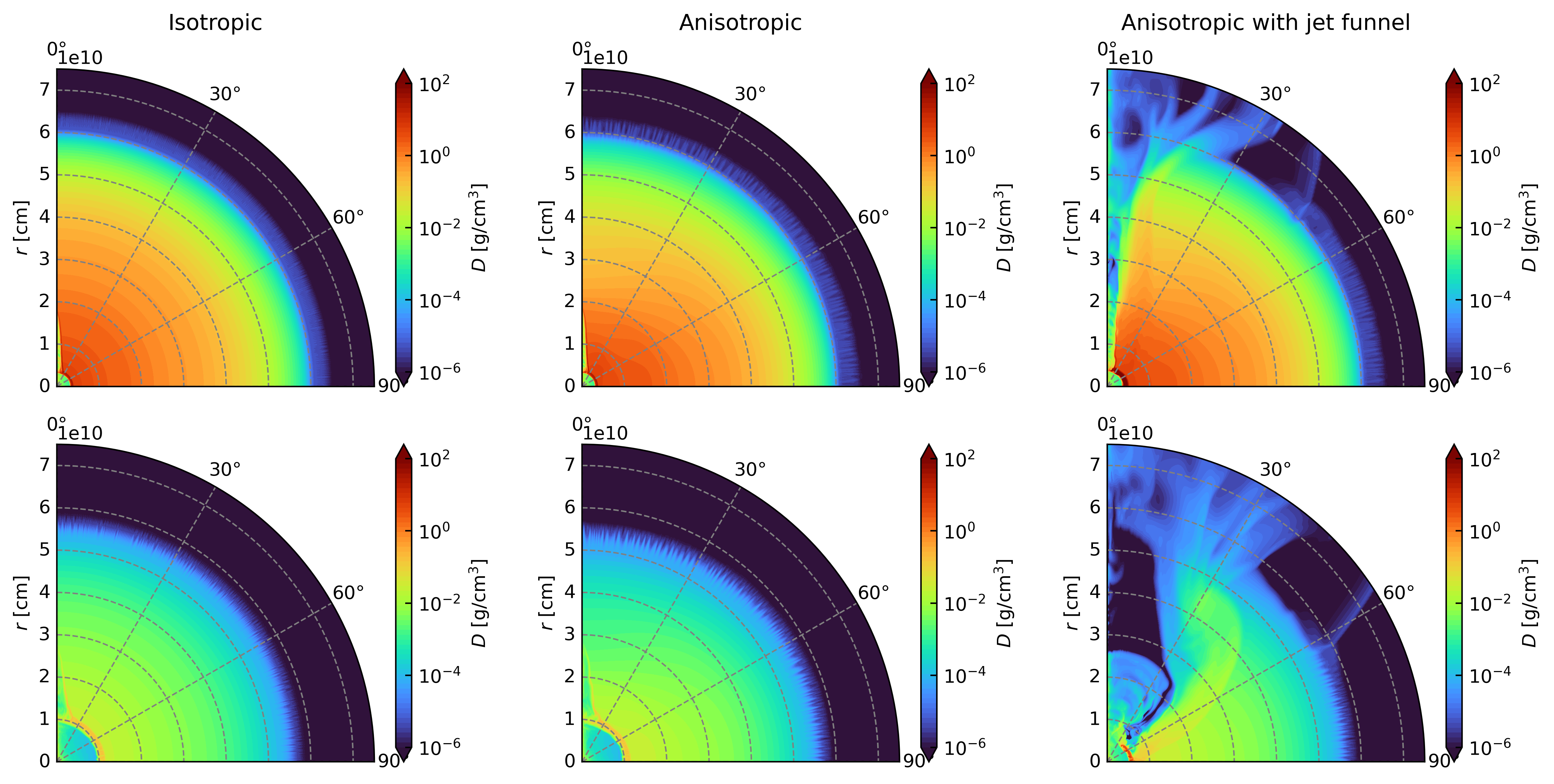}\\
\caption{Similar to Figure~\ref{fig:m-wind-rho}, but for Poynting-flux-dominated wind {with $\sigma=9$ }at the injection boundary. Upper panels: $M_{\rm ej}=10^{-1}M_\odot$ (model 8, 12  and 4); Bottom panels: $M_{\rm ej}=10^{-3}M_\odot$ (model 6, 10, 2).}
\label{fig:p-wind-rho}
\end{figure*}

\begin{figure*}
\includegraphics[width=2\columnwidth]{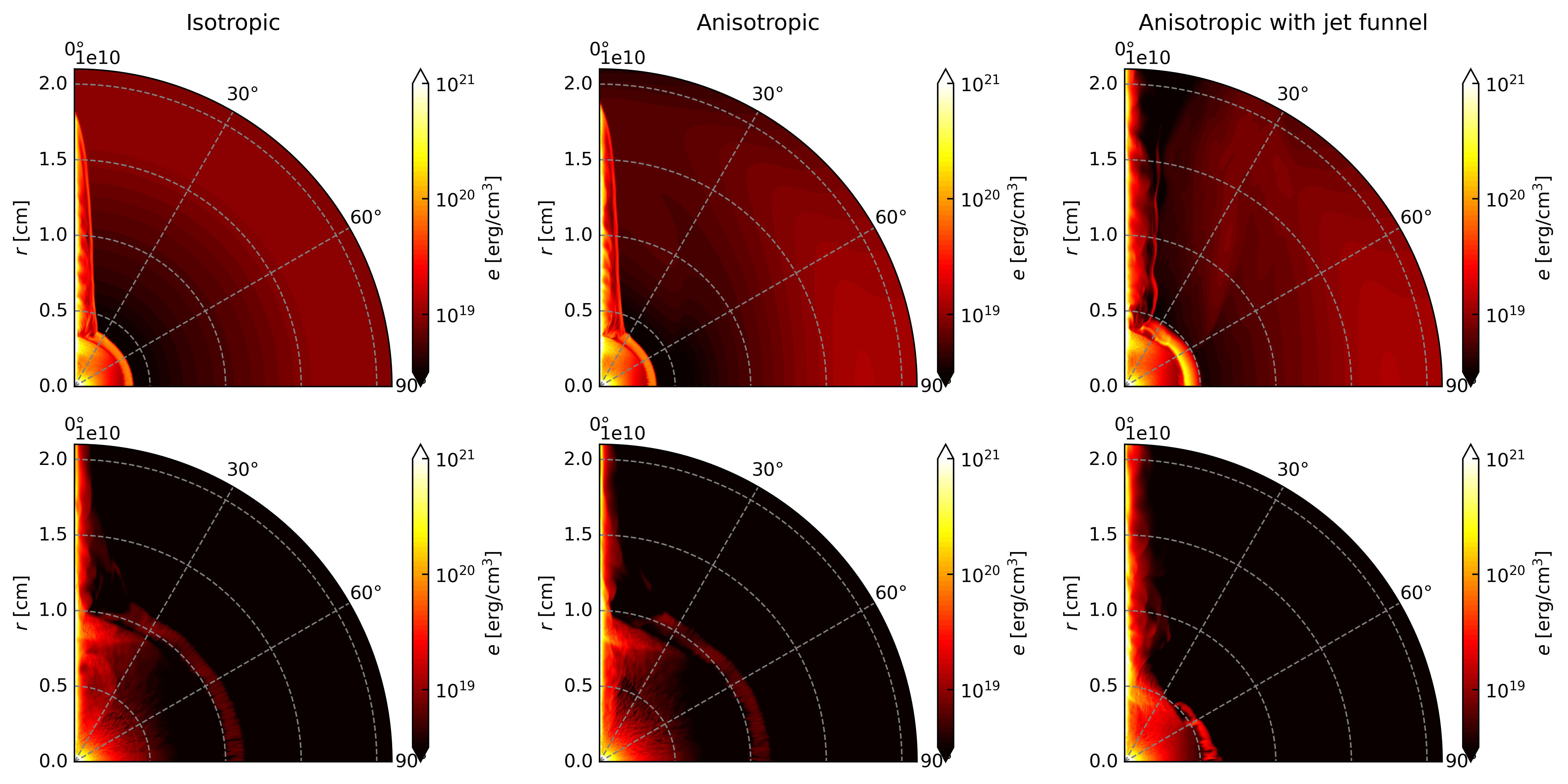}\\
\caption{Similar to Figure~\ref{fig:m-wind-e}, but for a Poynting-flux-dominated wind at the injection boundary. }
\label{fig:p-wind-e}
\end{figure*}

Similar to the findings discussed in Section~\ref{sec:m-wind}, Figure~\ref{fig:p-wind-rho} and Figure~\ref{fig:p-wind-e} provide snapshots of the density and energy density profiles at $t=5$ s for the Poynting-flux-dominated wind {($\sigma_w=9$)}. 
{As discussed in Section~\ref{sec:pure-wind}, if the magnetar wind is Poynting-flux-dominated before it collides with the ejecta, the resulting MWN will exhibit significant magnetization, resulting in a larger  $\langle \alpha_{\rm MWN}\rangle_{\rm E}$. The pinch effect from the hoop stress is stronger in comparison to the matter-dominated NWM, assuming that the NWM expands at the same speed. Consequently, this leads to the formation of more highly collimated outflows.

Upon comparing the left and middle panels of the figures, we can see that the additional collimation provided by the anisotropic ejecta has a negligible impact on the inherently high collimation of the wind. This holds true whether the ejecta is characterized by high density or low density. The primary source of collimation for the magnetar wind is the inherent MWN collimation mechanism.
}

\subsubsection{Injection energy distribution per solid angle}
We calculated the injection energy per solid angle for different wind and ejecta models by subtracting the total energy per solid angle at the end of the simulation ($t=15$ s) and the total energy per solid angle at the time when the wind starts to inject ($t=3$ s).

\begin{figure}
\includegraphics[width=\columnwidth]{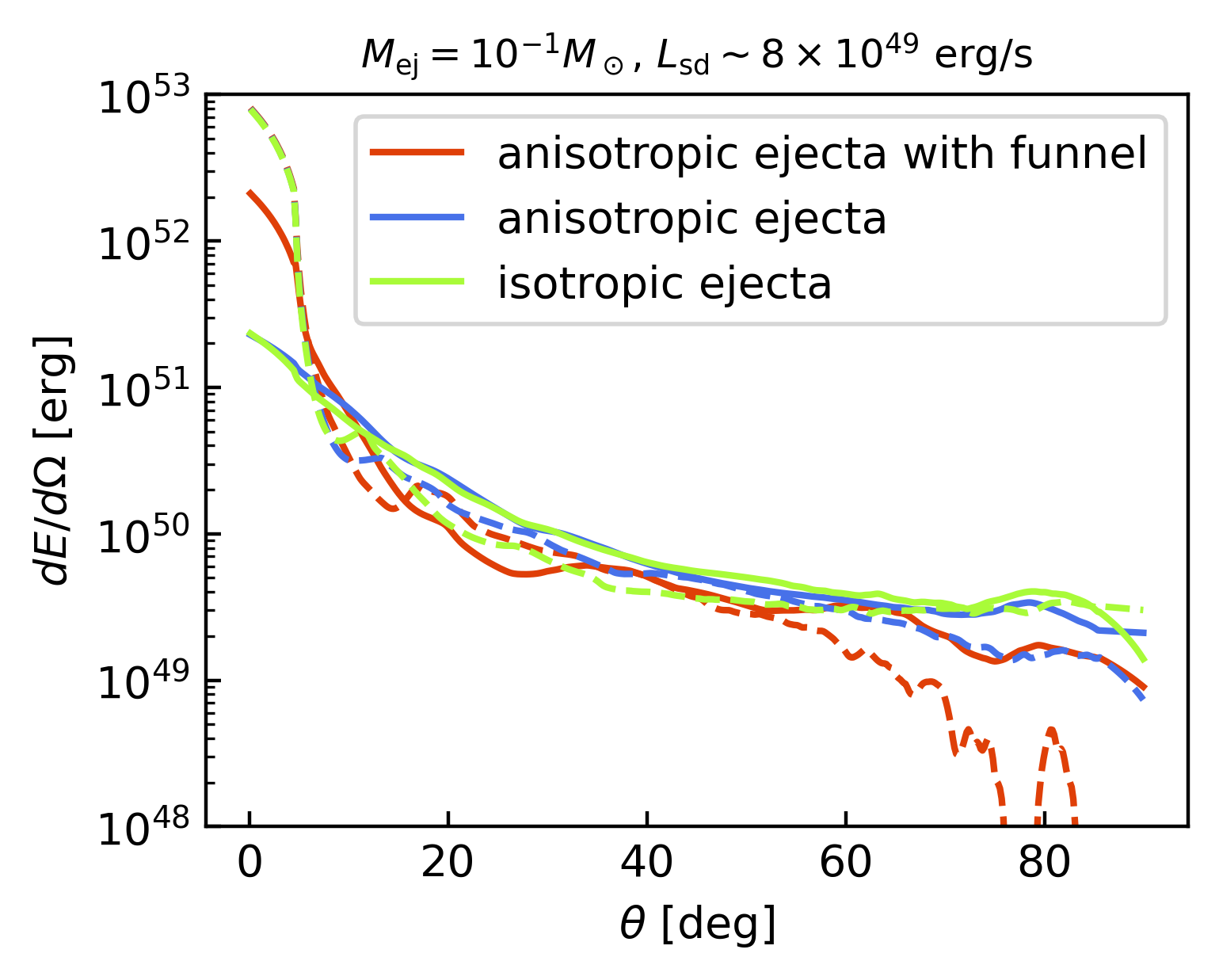}\\
\includegraphics[width=\columnwidth]{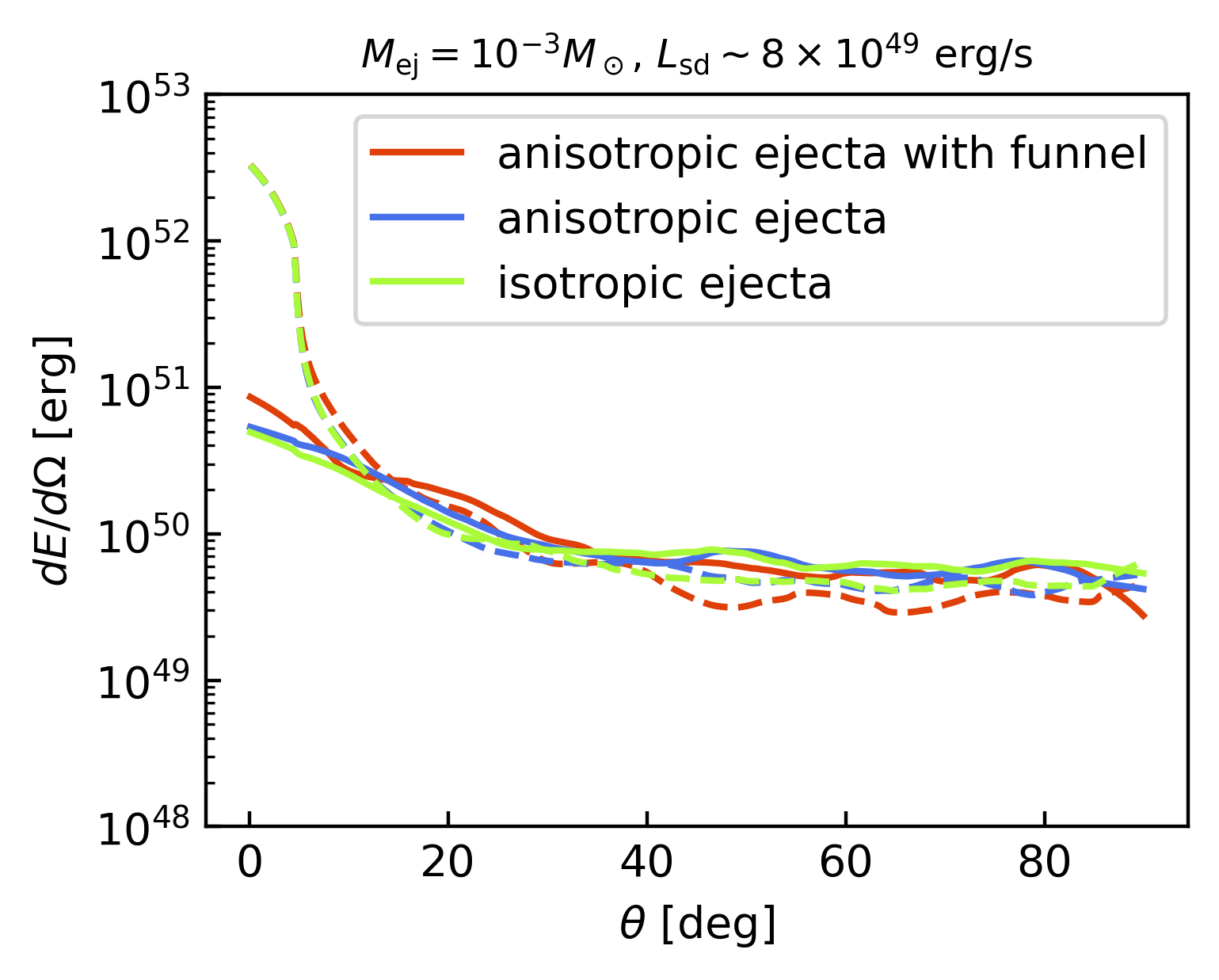}
\caption{Injected energy per solid angle for a magnetar wind.  The winding magnetic field is assumed to be monopole-like. The solid lines are for a matter-dominated wind with $\sigma=0.1$ and the dashed lines are for a Poynting-flux-dominated wind with $\sigma=9$. }
\label{fig:inject-e}
\end{figure}
Figure~\ref{fig:inject-e} presents the injection energy per solid angle for both matter-dominated wind (solid lines) and Poynting-flux-dominated wind (dashed lines). In the upper panels, where the wind's spin-down luminosity is approximately $10^{49}$ erg/s, and the lower panels, where it is roughly one order of magnitude lower, we observe distinct characteristics.

As discussed in Sections~\ref{sec:wind} and \ref{sec:m-wind}, for matter-dominated wind, collimation occurs through the interaction between the wind and the ejecta. In the upper panels of Figure~\ref{fig:inject-e}, where no pre-existing funnel is present, the wind injection appears to be insensitive to the ejecta profile. The green and blue lines are nearly indistinguishable. However, when a pre-existing funnel exists in the ejecta (red lines), we observe more concentration of energy at the pole due to the unobstructed path provided by the funnel.

In the case of Poynting-flux-dominated wind shown as dashed lines, the hoop stress in the non-relativistic magnetized MWN is exceptionally stronger, rendering the impact of the ejecta profile negligible in that region. Hence, the dashed lines appear identical near the pole, regardless of the ejecta profile being interacted with. 

In the case of low-density ejecta with a funnel shown in the bottom panel of Figure~\ref{fig:inject-e}, we can see that for both a matter-dominated and a Poynting-flux-dominated wind, the collimation becomes more insensitive to the ejecta profiles. The existence of the funnel adds a very weak extra collimation only to the matter-dominated wind.

It is evident that for both matter-dominated wind and Poynting-flux-dominated wind, the majority of the injected energy is collimated toward the pole rather than the equator for real sGRBs. As the magnetar's spin-down continues, with the spin-down luminosity decreasing as $L_{\rm sd}\sim t^{-2}$, there is a possibility that at very late stages, the wind will become trapped within the ejecta and heat it up. However, during the early stage, regardless of the wind model or ejecta model, the collimated wind can break out of the ejecta (depending on the total rotational energy budget and ejecta mass) and carve out a funnel. Therefore, even in the later stages of low-luminosity magnetar winds, they can still flow out through the funnel, making it unlikely for the wind to become trapped behind the ejecta.

\subsubsection{Wind injection efficiency}
The energy injection efficiency, particularly the heating efficiency, is crucial for engine-fed kilonovae since it determines the brightness of the resulting kilonova that can be observed directly. Previous studies have examined the energy injection efficiency in a 1D injection model. In this study, we calculate the energy injection efficiency for the 2D case. The energy injection efficiencies are calculated using the following equation:
\begin{eqnarray}
\xi=\frac{\int_{r_{\rm in}}^{r_{\rm out}}\mathcal{E}dV }{L_{\rm w}t}\,,
\end{eqnarray}
where $L_{\rm sd, iso}$ is the isotropic spin-down luminosity. For kinetic energy {  density} and thermal energy { density}, the expressions are as follows:
{ 
\begin{eqnarray}
\mathcal{E}_{\rm kin}&=&(\gamma-1)Dc^2,\\
\mathcal{E}_{\rm therm}&=&\gamma^2\frac{\Gamma}{\Gamma-1}p_g - p_g,
\end{eqnarray}
}
respectively. The total injection efficiency is $\xi_{\rm tot} = \xi_{\rm kin}+\xi_{\rm therm}$.

\begin{figure*}
\includegraphics[width=2\columnwidth]{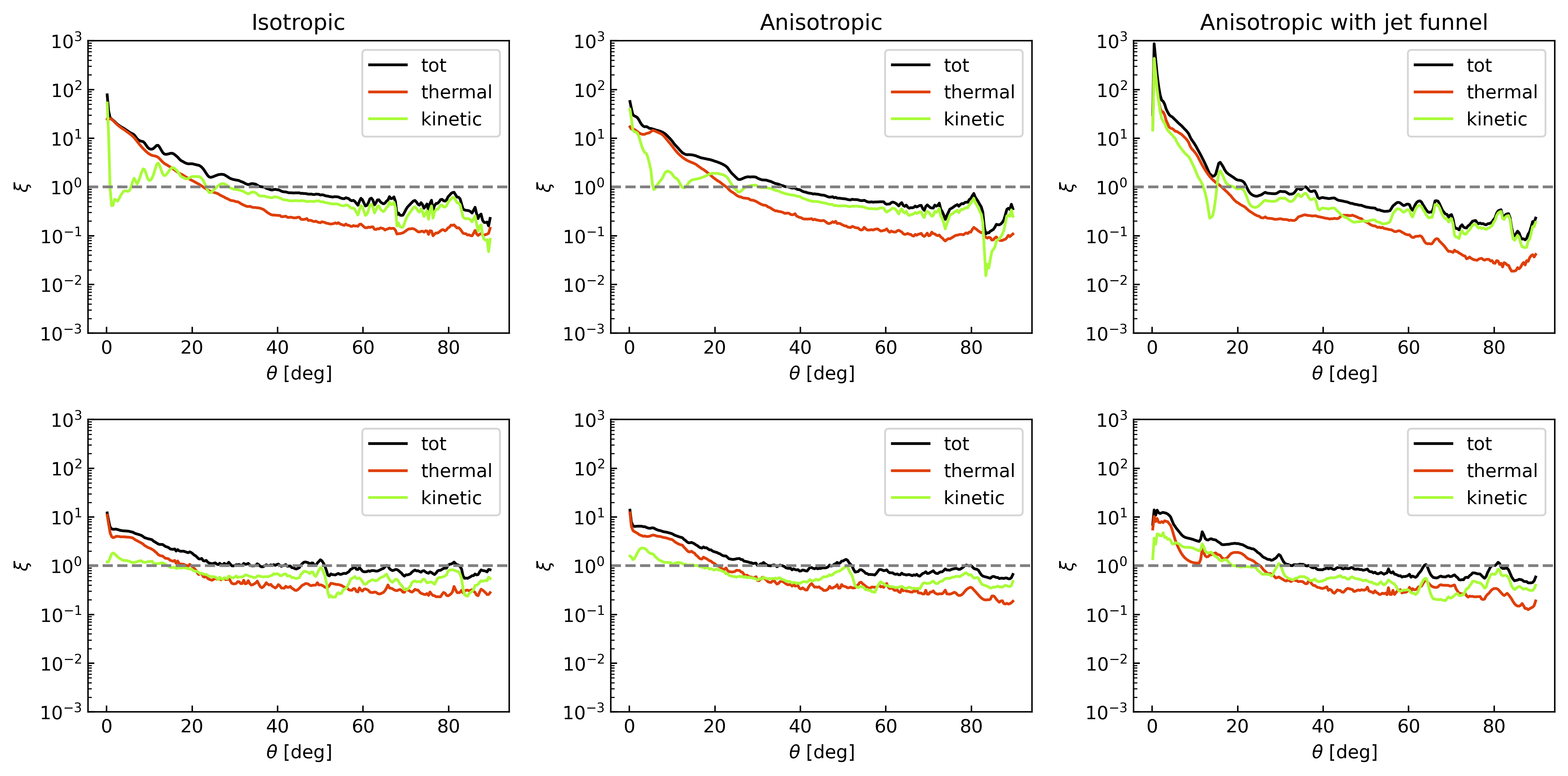}
\caption{Matter-dominated ($\sigma_{\rm w}=0.1$) wind injection efficiency for different ejecta models. Upper panels: total ejecta mass is $10^{-1}M_\odot$; Bottom panels: total ejecta mass is $10^{-3}M_\odot$. }
\label{fig:xi-m}
\end{figure*}

Figure~\ref{fig:xi-m} illustrates the matter-dominated isotropic wind injection efficiency for various ejecta profiles. The upper three panels correspond to a total ejecta mass of $10^{-1}M_\odot$, while the bottom three panels represent $10^{-3}M_\odot$.

In the case of high-density ejecta, the total energy injection efficiency ranges from approximately {20\% to 30\%} at the equator and reaches values as high as roughly {$10^{4}$\%} near the pole. The majority of the injected energy is converted into kinetic energy, as indicated by the green lines. For collimated outflow near the pole, the dominance of kinetic energy over thermal energy suggests that the outflow is accelerated to high speeds. The magnetic energy does not peak at the pole due to the $\sin^2\theta$ scaling of the injected magnetic field. In this scenario, the collimation is not strong enough to focus an adequate amount of magnetic energy toward the pole. For ejecta profiles featuring a pre-existing funnel, the energy injection efficiency near the pole can be as high as 1000, owing to the narrowness of the pre-existing funnel.

In the case of low-density ejecta ($M_{\rm ej}=10^{-3}M_\odot$), the wind injection is more isotropic. The total energy injection efficiency is approximately {60\%} to {100\%} near the equator and around {$10^3$\%} near the pole. Similar to the high-density ejecta case, most of the energy is converted into kinetic energy. If no pre-existing funnel is present, the kinetic energy exhibits a nearly isotropic behavior. Due to the weak confinement in this situation, thermal acceleration is inefficient. Consequently, at the pole, the outflow is predominantly thermally dominated.

\begin{figure*}
\includegraphics[width=2\columnwidth]{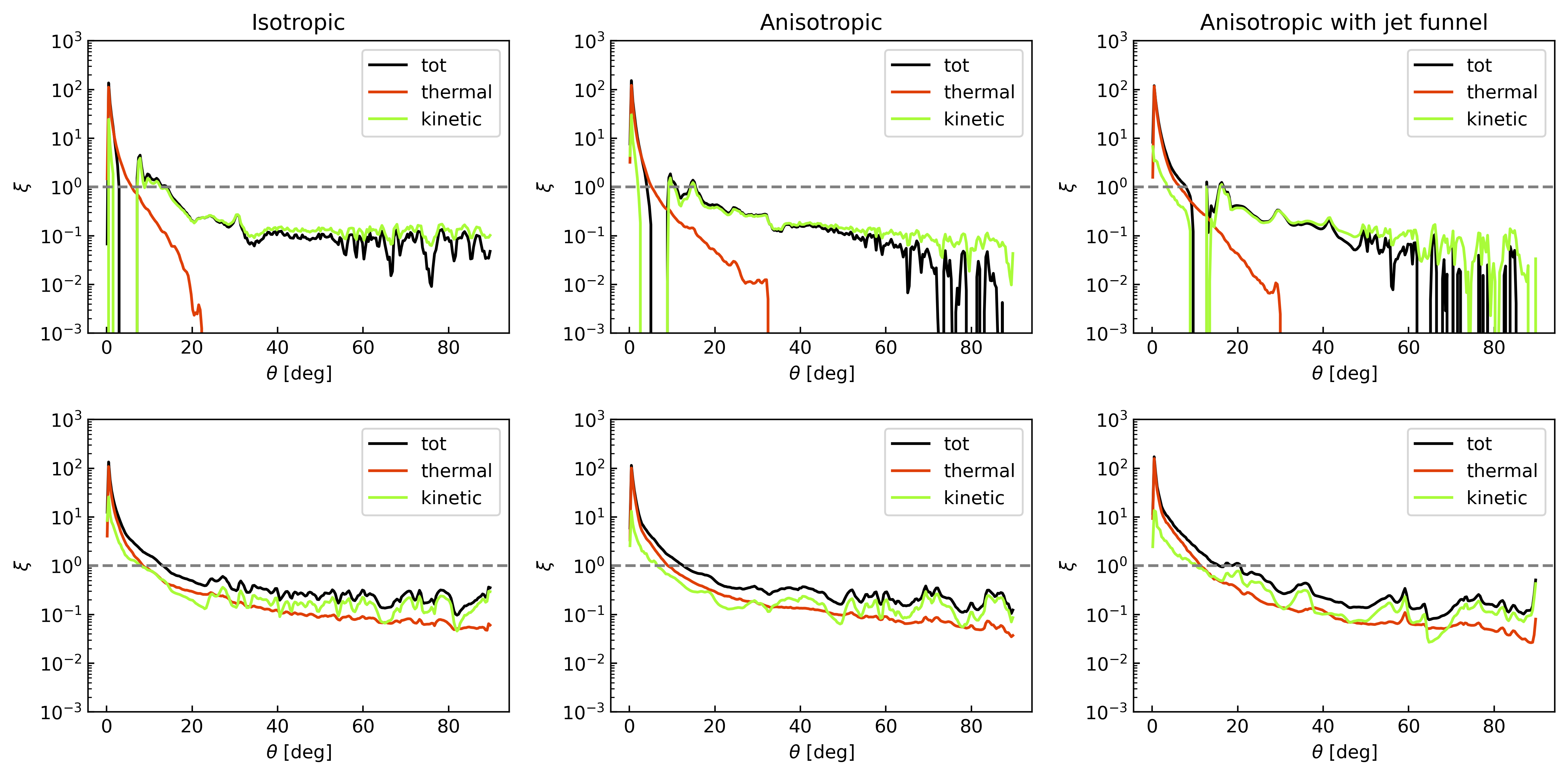}
\caption{ Similar to Figure~\ref{fig:xi-m}, but for Poynting-flux-dominated ($\sigma_{\rm w}=9$) wind injection.}
\label{fig:xi-p}
\end{figure*}
Figure~\ref{fig:xi-p} depicts the wind injection efficiency for Poynting-flux dominated scenarios. Due to strong collimations, the total energy injection efficiency near the equator is significantly lower compared to matter-dominated winds. The injection efficiency near the equator is below {10\%}, while it reaches values as high as {$10^{4}$\%} near the pole for $M_{\rm ej}=10^{-1}M_\odot$. Similar to matter-dominated winds, most of the injected energy is converted into kinetic energy within the ejecta. However, for the outflow near the pole, the plasma is dominated by magnetic energy. This is attributed to persistent energy injection and strong magnetic collimation. Consequently, even though thermal/magnetic acceleration is involved, magnetic energy remains dominant in this region.

It is worth noting that the thermal energy injection efficiency is negative for $\theta>20-30^\circ$, which may not be clearly visible on a logarithmic scale. This negative thermal efficiency arises due to the inclusion of ejecta expansion in our efficiency calculation. The heat flow from the equator to the pole within the ejecta contributes to the negative thermal efficiency near the equator. For low-density ejecta ($M_{\rm ej}=10^{-3}M_\odot$), the wind is less efficient than in the high-density ejecta case for accelerating the ejecta, and the heat flow from the equator to the pole is also weaker.\\

{
\citet{Ai2022} investigated the 1-D wind injection case. In their work, they assumed an isotropic Poynting-flux dominant wind ($\sigma_w=10^4$, $\gamma_w=10^3$) with spin-down luminosity $L_{\rm}$ ranges from $10^{45}$-$10^{49}$ erg/s. In their calculations, they found that a significant fraction of magnetic energy can be converted to kinetic energy by wind-induced ejecta accelerations. For $10^{-3}M_\odot$ ejecta, 70\% of the injected energy will be stored as magnetic energy and kinetic energy, and less than 30\% injected energy can be converted to internal energy via reverse shock heating. 
To compare with their result, we calculate the total injection efficiency as well. For Poynting-flux dominated wind ($\sigma=9$, $L_{\rm sd}\sim 10^{49}$ erg/s) with $10^3M_\odot$ in our simulations, we also find that roughly $70\%$ of the injected energy is stored as magnetic energy and kinetic energy. To be more specific, roughly 35\% of the energy is stored as magnetic energy while roughly 35\% of the injected energy is converted into kinetic energy, regardless of the pre-injection ejecta profiles. As the spin-down luminosity drops to $10^{48}$ erg/s, the kinetic energy conversion efficiency drops to 30\%, while the total fraction of magnetic energy and kinetic energy maintains around 70\%. All those findings are consistent with the results in \citep{Ai2022}. {  The heating efficiencies of different models are summarized in Table~\ref{tab1} .}
}

\section{Conclusions and Discussion}

\subsection{Conclusions}
This paper presents investigations of the wind injection process from a magnetar in sGRBs using numerical simulations based on special relativistic magnetohydrodynamics (SRMHD). The study aims to understand the criteria for collimated outflow to appear, efficiency and angular distribution of energy injection into the surrounding medium. Our conclusions are summarized below.

\begin{itemize}
    \item Most of the magnetar wind injection is non-isotropic. The asymmetry of the wind injected is determined by the magnetization and velocity of the expanding magnetar wind nebula that formed from wind ejection collision. High magnetization and low expanding speed result in the collimated outflow. The criteria $u_{\rm A}/u_{\rm MWN} \gg 1$ is a good indicator for wind collimation.

    \item Isotropic magnetar wind injection is hard to achieve. This requires either a very low magnetization in the MWN or an ultra-relativistic moving MWN. The low magnetization of MWN needs the magnetar wind to be accelerated efficiently in the wind acceleration region with strong magnetic dissipation, such that the magnetization of the wind can be decreased far below 1. The ultra-relativistically moving MNW requires the total injected energy from the magnetar wind to be much higher than the ejecta rest mass. 
    
    \item For the collimated injection process, the anisotropy of the wind injection is insensitive to the ejecta profiles. However, a pre-existing funnel created by the GRB jet helps with the matter-dominated wind blowing out from the pole.

    \item If magnetars are proposed as the engine for an engine-fed kilonova, the energy injection efficiency to the ejecta might be lower than what we expected before, due to the wind collimation. { The typical energy injected, which is generally estimated from the total spin-down energy of a magnetar per solid angle will be smaller by a factor of up to 10 with respect to the value of the isotropic case due to wind collimation}. 

\end{itemize}

\subsection{Discussion}

Our results have implications for understanding the sGRB phenomenology. There are three possible criteria to diagnose the existence of a magnetar engine in sGRBs. The first criterion is the extended emission or internal X-ray plateau as observed in a good fraction of sGRBs \citep[e.g.][]{Rowlinson2013,Lv2015}. The typical energy emitted in the X-ray plateau phase is usually much smaller than the total spin energy of a millisecond magnetar. This has been attributed to a possible low X-ray emission efficiency and possible energy loss due to secular gravitational wave emission of the post-merger magnetar \citep[e.g.][]{Fan2013,Gao2016,Xiao2019}. Our results suggest that the injected energy { estimated from the total spin-down energy of the magnetar} into the polar region that powers the X-ray plateau emission may be even larger, which demands an even lower radiation efficiency or an even larger gravitational wave loss. 

The second criterion is the brightness of the sGRB associated kilonova. Some recent studies \citep[e.g.][]{Wang2023} have used the non-detection of bright kilonovae (but see \citealt{Gao2017}) to argue that long-lived magnetar engine in binary neutron star mergers are rather rare. They have assumed an isotropic energy injection and a relatively large heating efficiency to reach the conclusion. Our results suggest that energy injection to kilonova is smaller by a factor of up to 10 than previously assumed for the same magnetar parameters. This result, together with the finding that the heating efficiency is quite low \citep{Ai2022}, suggest that the current result could be consistent with the hypothesis that a significant fraction of binary neutron star mergers leaves behind a magnetar engine. The emergence of a magnetar engine from the recently detected long-duration merger-type (Typpe I) GRB 230307A, which has a not-too-bright kilonova \citep{levan2023,yang2023}, suggest that it is entirely possible to have a magnetar-fed kilonvoa that appears as a regular kilonova detected in many sources.

The final criterion to test a magnetar engine is via the late-time radio afterglow energy of sGRBs \citep{Metzger2014b,Fong2016,Horesh2016} or the he radio afterglow of the non-relativistic kilonova ejecta \citep{Nakar2011,Gao2013,Hotokezaka2015}. At late time when the jet becomes non-relativistic, the total injected energy { from the magnetar spin-down} to essentially all the directions will be probed. Since our results only re-distribute the energy in different angles, it would not modify the late-time afterglow radio flux significantly. By properly accounting for late-time dynamics, \cite{Liu2020} showed that the magnetar engine is allowed for most sGRBs in a large parameter space. If one focuses on the kilonova ejecta emission only, then the predicted flux should be weaker than previously predicted assuming isotropic energy injection. 

\section*{Acknowledgements}
We acknowledge NASA award 80NSSC23M0104 and the Nevada Center for Astrophysics for support. ZZ also acknowledges support by NASA award 80NSSC22K1413. YW acknowledges helpful discussions with Jinjun Geng during the early stages of this paper.

\section*{Data Availability}
The data underlying this article will be shared on reasonable request to the corresponding author.



\bibliographystyle{mnras}
\bibliography{example} 



\appendix
\onecolumn
\section{Forces on fluid elements}
To gain a better understanding of the collimation effect described in Section~\ref{sec:pure-wind}, this Appendix presents calculations of the forces acting on each fluid element, illustrating their relative strengths under various conditions. In the context of relativistic MHD, the momentum equation being solved can be expressed as
\begin{eqnarray}
\frac{\partial \vec{S}_{\rm tot}}{\partial t} + \nabla \cdot [\vec{S}_{\rm tot}\vec{v}+p_{\rm tot}\mathbf{I} -\mathbf{E}\mathbf{E}-\mathbf{B}\mathbf{B}]=0
\end{eqnarray}
This equation can be re-written as
\begin{eqnarray}
\frac{D \vec{S}_{\rm tot}}{D t}&=&\frac{\partial \vec{S}_{\rm tot}}{\partial t} + \nabla \cdot[\vec{S}_{\rm tot}\vec{v}]=-\nabla[p_g+\frac{1}{2}(B^2+E^2)] + \nabla\cdot(\mathbf{E}\mathbf{E}+\mathbf{B}\mathbf{B})\\
&=&-\nabla p_g -\frac{1}{2}\nabla B^2 + (\nabla\cdot \mathbf{B}) \mathbf{B} +\mathbf{B} \cdot \nabla \mathbf{B} - \frac{1}{2}\nabla E^2 + (\nabla\cdot \mathbf{E}) \mathbf{E} +\mathbf{E} \cdot \nabla \mathbf{E}\\
&=& -\nabla p_g + (\nabla\times\mathbf{B} )\times \mathbf{B} - \mathbf{E} \times(\nabla\times\mathbf{E}) + (\nabla\cdot\mathbf{E}) \mathbf{E} \label{eq:forces}
\end{eqnarray}
The first term in Eq~\ref{eq:forces} represents the fluid acceleration due to the gas pressure gradient, while the second term corresponds to the Lorentz force. The last two terms represent the Coulomb force. In MHD simulations, the electric field $\mathbf{E}$ can be substituted with $\mathbf{v}$ and $\mathbf{B}$ by applying the ideal MHD condition:
\begin{eqnarray}
\mathbf{E}+\frac{\mathbf{v}}{c}\times\mathbf{B} = 0\,.
\end{eqnarray}
Using this, we can evaluate the three components (gas pressure, Lorentz force, and Coulomb force) and compare their relative strengths under the different conditions discussed in Section~\ref{sec:pure-wind}.

\begin{figure}
\includegraphics[width=.5\columnwidth]{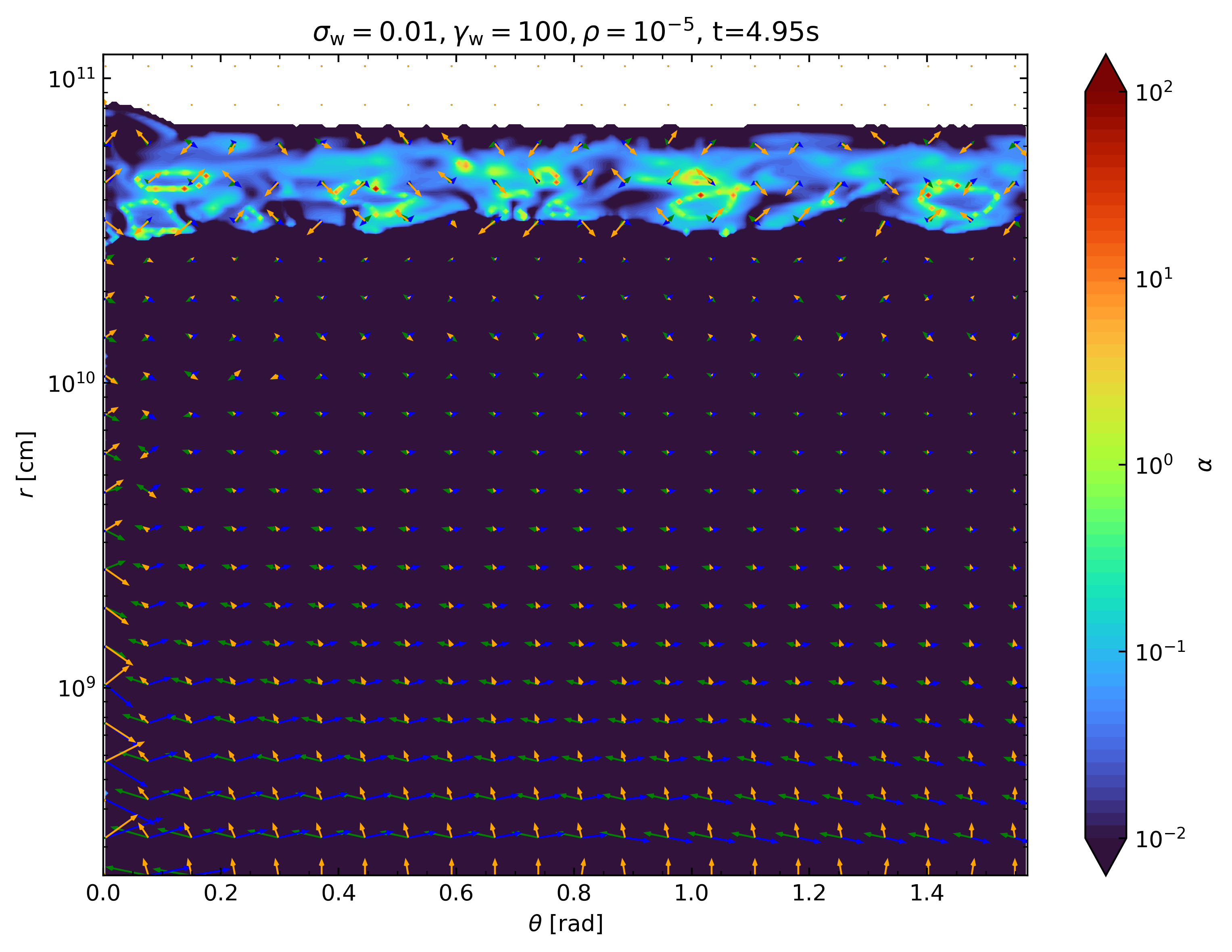}
\includegraphics[width=.5\columnwidth]{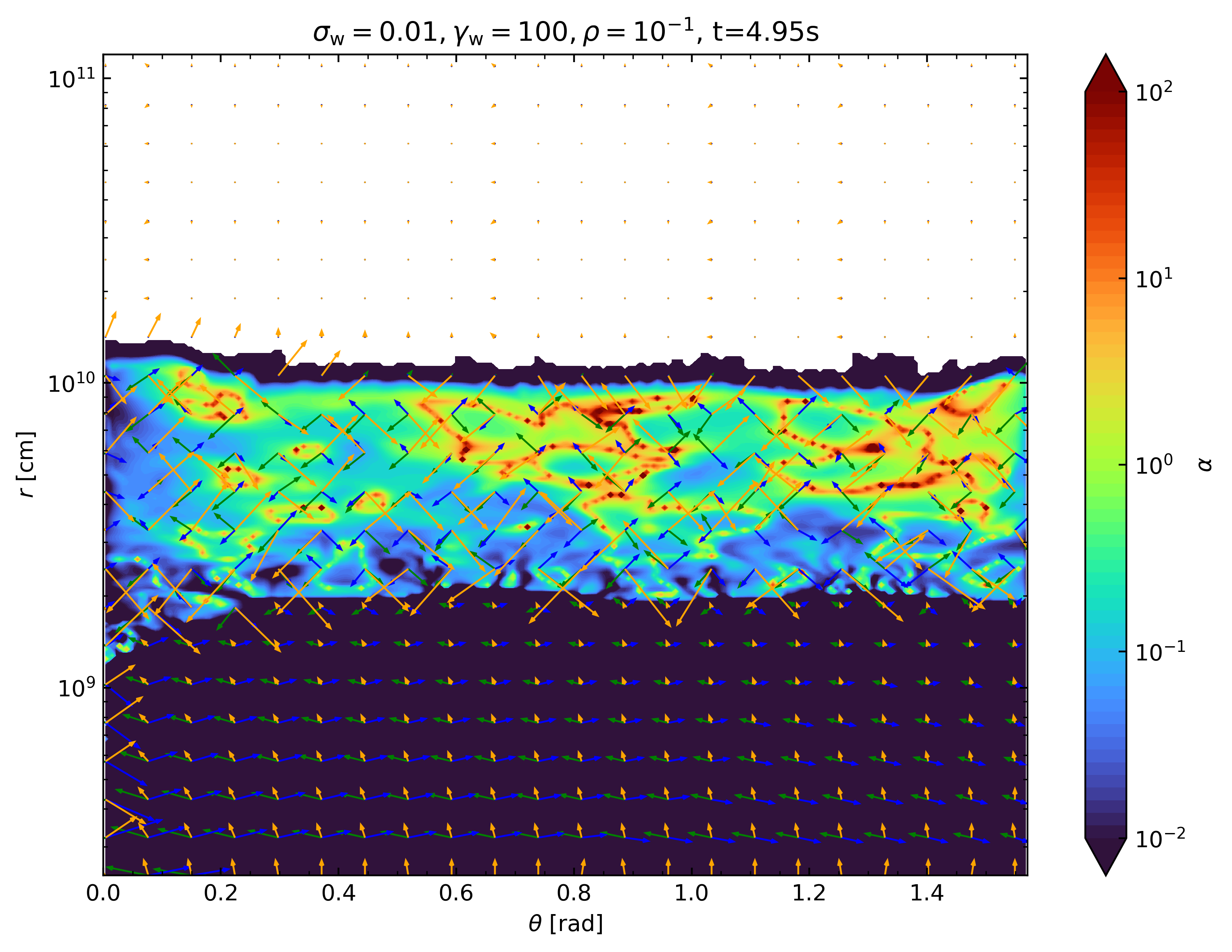}
\caption{Fluid element acceleration calculated by using Eq~\ref{eq:forces}. Blue vectors represent the Coulomb acceleration corresponding to the last two terms in Eq~\ref{eq:forces}; green vectors denote the Lorentz acceleration from the second term; and orange vectors illustrate the acceleration due to the gas pressure gradient. The lengths of these vectors are displayed on a logarithmic scale.}
\label{fig:f001}
\end{figure}

\begin{figure}
\includegraphics[width=.5\columnwidth]{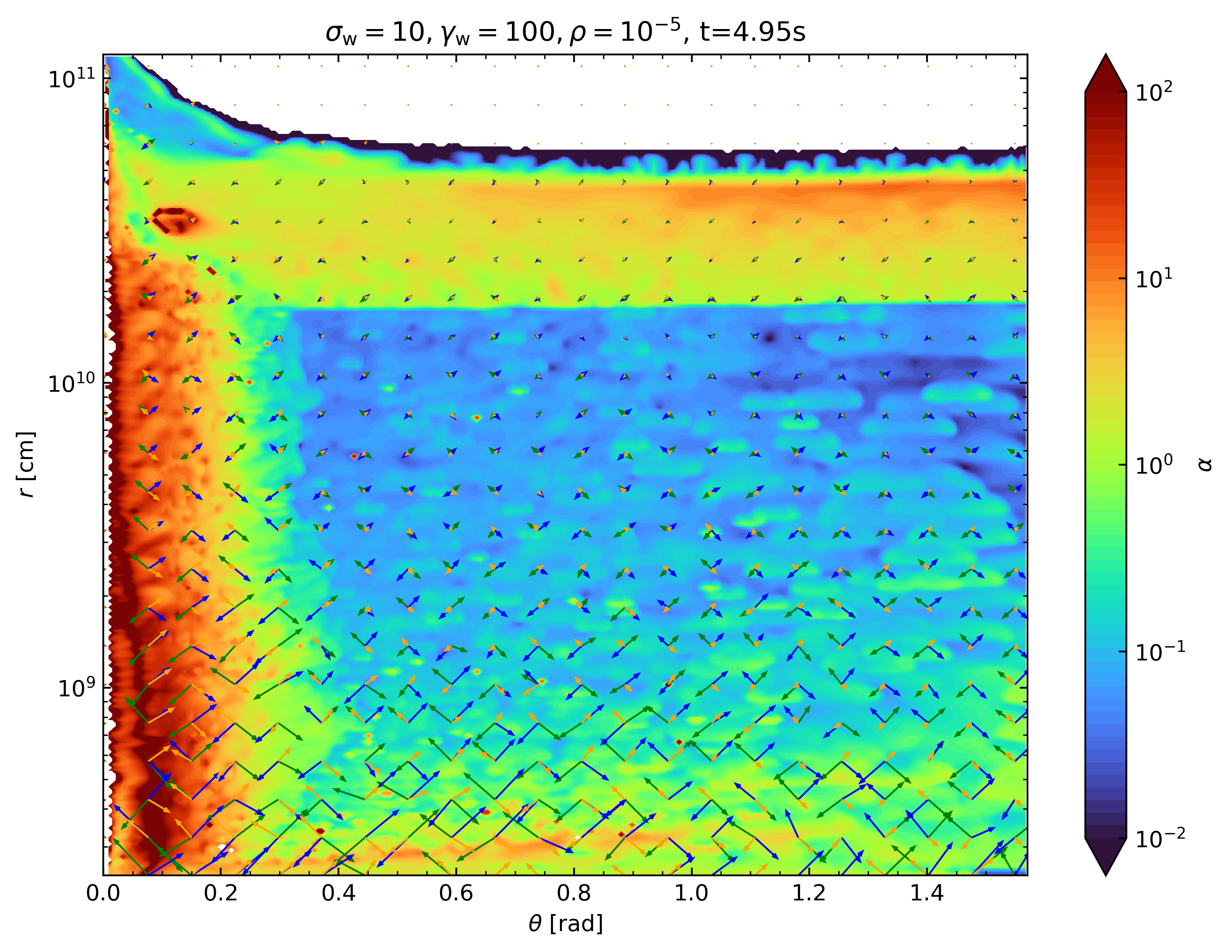}
\includegraphics[width=.5\columnwidth]{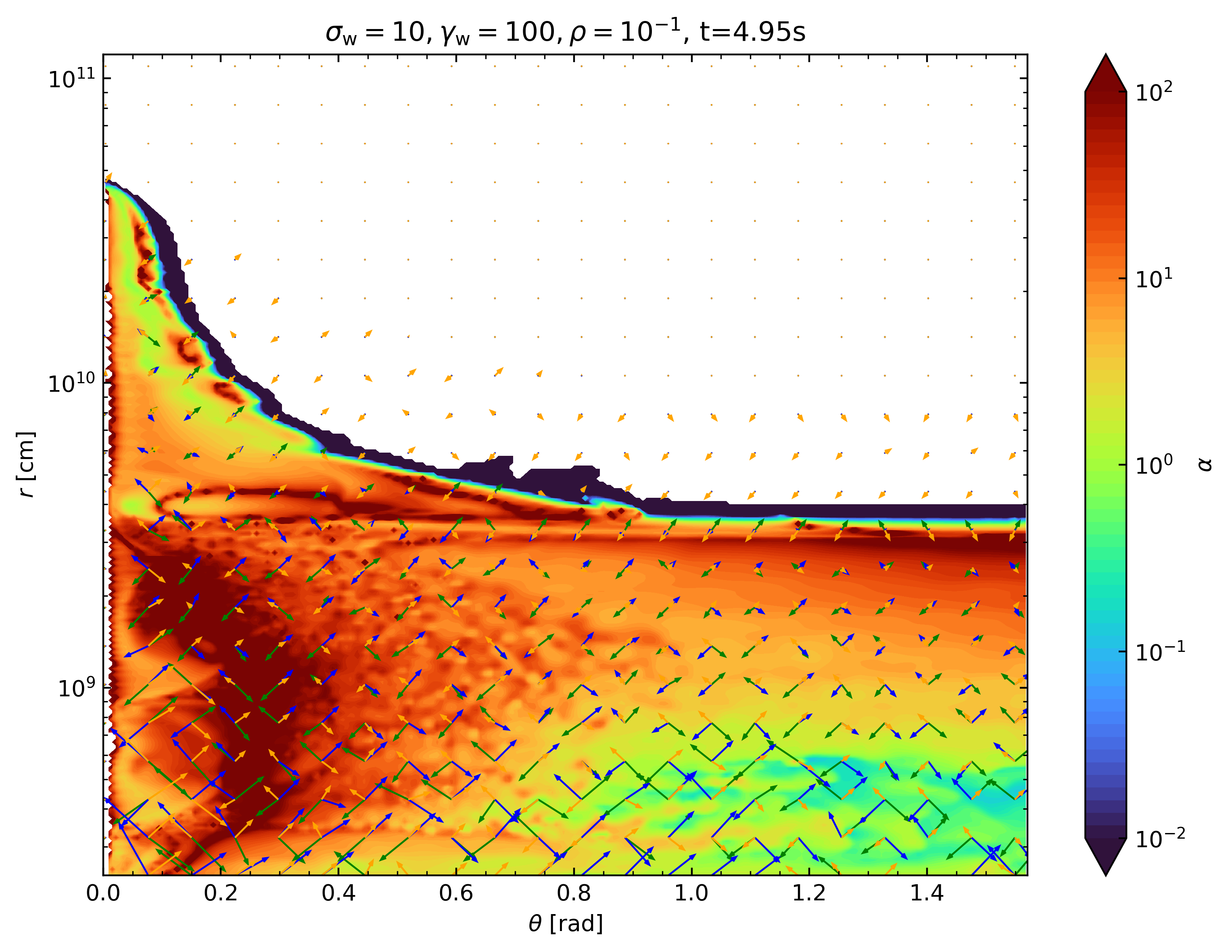}
\caption{Same as Fig~\ref{fig:f001}, but for higher initial wind magnetization $\sigma_w=5$.}
\label{fig:f10}
\end{figure}
Figure~\ref{fig:f001} and \ref{fig:f10} show the acceleration of fluid elements for different values of $\alpha$. It is evident that in low $\alpha$ regions, the Lorentz force and Coulomb force are well balanced, while in high $\alpha$ regions, they become unbalanced. The collimation arises from the high $\alpha$ regions.

\bsp	
\label{lastpage}
\end{document}